\documentclass[superscriptaddress,prl,aps,amssymb,twocolumn,10pt]{revtex4-1}

\usepackage{amsmath}
\usepackage{amssymb}
\usepackage{amsthm}
\usepackage{amsfonts}
\usepackage{listings}
\usepackage{enumerate}
\usepackage{latexsym}

\usepackage[mathscr]{euscript} 

\usepackage[colorlinks=true,citecolor=blue,linkcolor=blue,urlcolor=black]{hyperref} 

\usepackage{wasysym}

\usepackage{psfrag}

\usepackage{bm}

\usepackage{bbm} 

\usepackage{graphicx}

\usepackage{color}

\usepackage{braket}

\newcommand{\beq}{\begin{equation}}
\newcommand{\eneq}{\end{equation}}

\newcommand{\dd}{\mathrm{d}}

\def\ie{{\it i.e.},\ }
\def\eg{{\it e.g.},\ }

\newcommand{\e}{\mathrm{e}}
\newcommand{\kk}{{\bf k}}
\newcommand{\RR}{{\bf R}}
\newcommand{\GG}{{\bf G}}
\newcommand{\rr}{{\bf r}}
\newcommand{\K}{{\bf K}}

\newcommand{\diff}{{\mathrm{d}}}

\newcommand{\Li}{{\lambda_i}}

\begin{document}

\title{Embedded Topological Insulators}
\date{\today}
\author{Thomas I. Tuegel}
\affiliation{Department of Physics, and Institute for Condensed Matter Theory, University of Illinois at Urbana-Champaign, Urbana, IL 61801}
\author{Victor Chua}
\affiliation{Department of Physics, University of Basel, Klingelbergstrasse 82, CH-4056 Basel, Switzerland}
\author{Taylor L. Hughes}
\affiliation{Department of Physics, and Institute for Condensed Matter Theory, University of Illinois at Urbana-Champaign, Urbana, IL 61801}

\begin{abstract}

We present a generalization of free fermionic topological insulators that are composed of topological subsystems of differing dimensionality. These topological subsystems of nonzero co-dimension are embedded within a trivial insulating environment. A general procedure is described to isolate and classify such embedded topological insulators and we present three representative examples in varying dimensions and symmetry classes. Moreover, we demonstrate with concrete examples that the presence of periodically placed embedded topological insulators in an otherwise trivially classified system can lead to topologically non-trivial physical phenomena on crystalline defects; namely, novel topological surface/edge modes at stacking faults and partial edge dislocations.


\end{abstract}


\maketitle


After a decade of intense effort, the classification of free-fermion topological insulators and superconductors has been thoroughly developed. Beginning with the strong topological phases (STIs) that are protected only by time-reversal, charge-conjugation, and/or chiral symmetries~\cite{kane_$z_2$_2005,kane_quantum_2005,bernevig_quantum_2006,schnyder_classification_2008,kitaev_periodic_2009,qi_time-reversal-invariant_2009,ryu_topological_2010,qi_topological_2011}, the scientific focus has moved on to weak(er) topological phases protected by lattice translation~\cite{fu_topological_2007,moore_topological_2007,roy_topological_2009} and point-group symmetries~\cite{fu_topological_2011,benalcazar_classification_2014}. An important outgrowth of the basic theory was the understanding of the interplay between symmetry, topology, and defects. This has led to a rich set of phenomena associated to the generation of stable fermionic bound states localized on various types of topological defects, \eg vortices, dislocations, or disclinations~\cite{ran_one-dimensional_2009,teo_topological_2010}.



Recall that the combination of two non-interacting topological insulators (TIs) of same symmetry class and \emph{equal dimensionality} is classified by the sum of their topological invariants.\footnote{In fact the method of topological classification using topological K-theory~\cite{kitaev_periodic_2009,teo_topological_2010,freed2013twisted,shiozaki2014topology} is precisely meant to respect this sum rule by determining the appropriate functor from equivalence classes of smooth vector bundles to suitable algebraic rings.}
This remains true even if the constituent topological subsystems are coupled and/or if weak disorder is present, so long as the bulk energy gap remains robust, and the protective symmetries are unbroken.
In this article, we pose and answer a simple question:
What then happens if two insulators (topological or trivial) of \emph{different dimensionalities} but common symmetry class are combined and coupled? We partially address this question in the case when one subsystem is a trivial insulator by demonstrating that such a composite system can retain its topologically protected surface states in a finite region of stability. This then leads us to the notion of Embedded Topological Insulators (ETIs) which are topological insulators that are embedded -- sometimes as defects -- within a trivial insulating environment of greater dimensionality.
While in some cases this notion is closely related to the weak topological insulators (WTIs), it is notably distinct.
Indeed, some natural situations that we envisage are insulators that are generally inhomogeneous in space, \eg due to defects.

\begin{figure}
\includegraphics[width=8.5cm]{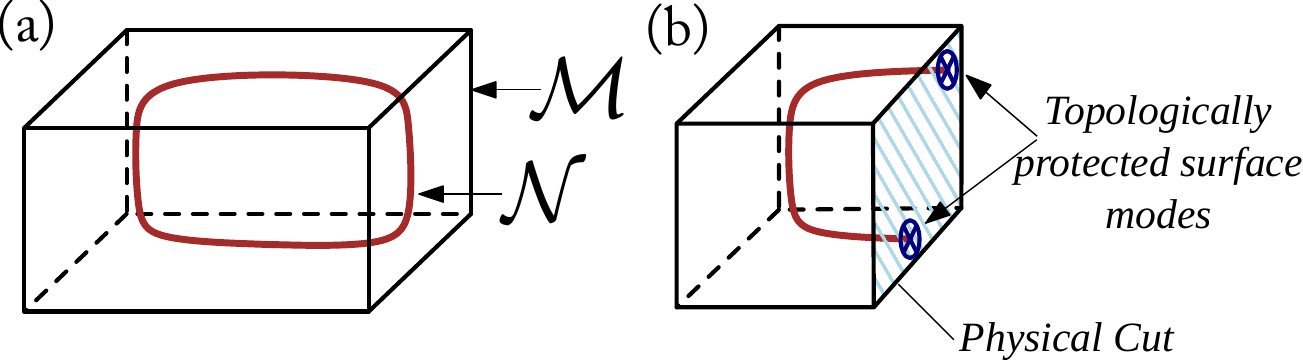}
\caption{Example of a composite insulator system with two thermodynamically large insulator subsystems ($\mathcal{M}$ and $\mathcal{N}$) of different dimensionalities. (a) A closed system with a trivial 3D environment $\mathcal{M}$, and a closed topological wire $\mathcal{N}$ that is the isolated Embedded Topological Insulator (ETI). (b) The topological surface states which are localized bound states are exposed whenever the composite system is cleaved such that the physical surfaces cuts through $\mathcal{N}$.
}
\label{fig:schematic}
\end{figure}

We shall explore first the case of only two constituent insulators of differing dimensions.
We denote the subsystem with greater dimension the environment $\mathcal{M},$
and the embedded subsystem with smaller dimension as $\mathcal{N}$.
The subsystems $\mathcal{M}$ and $\mathcal{N}$ are assumed to be thermodynamically large in their respective dimensions. The interesting case corresponds to $\mathcal{N}$ being topologically non-trivial in its `intrinsic' dimension while $\mathcal{M}$ is trivial such that $\mathcal{N}$ is the ETI. An example of the situation that we have just described is shown in Fig.~\ref{fig:schematic}(a). Later, we generalize our discussion to an arbitrary number of ETIs within a common trivial environment. The converse situation where $\mathcal{M}$ is topologically non-trivial and $\mathcal{N}$ trivial is not as interesting since $\mathcal{N}$ essentially behaves as an impurity which trivially disorders the total system, although it may experience induced topological proximity effects\cite{hsieh_bulk_2016}.
The more complicated case when both $\mathcal{M}$ and $\mathcal{N}$ are topological will not be discussed here but warrants future study.

\paragraph*{Topological Invariants}
The key to computing invariants for any translation-invariant TI
is the local (i.e., defined at each momentum point) spectral projector $P(\kk)$
\footnote{See the supplementary materials for an introductory discussion.}
\begin{equation}
P(\kk) = \sum_{E_{n\kk}< E_F} |u_n(\kk)\rangle \langle u_n(\kk)|
\end{equation}
where $E_F$ is the Fermi energy
\footnote{In the cases of particle-hole symmetry, we tune the Fermi energy to the particle-hole symmetric point conventionally set at $E_F=0$.}
and $\{ E_{n\kk}\},\{|u_n(\kk)\rangle\}$ are the energy bands and periodic Bloch functions respectively.
The integral over the Brillouin zone (normalized by the reciprocal unit cell volume $|C^*|$)
of the local projector gives the total projector,
\begin{equation}
P = \int_\text{BZ}\frac{\dd^d k}{|\mathcal{C}^*|} P(\kk) \otimes |\kk \rangle \langle \kk|,
\end{equation}
which respects any symmetry $g$ of the 1-body Hamiltonian $H.$
The strong topological invariants
\cite{kitaev_periodic_2009,schnyder_classification_2008,qi_topological_2008}
depend on the symmetry class and dimension $d,$ and characterize the smooth vector bundle defined by the local spectral projector $P(\kk)$\cite{chiu2016classification,shiozaki2014topology}.

We limit ourselves to studying time-reversal ($\Theta$), charge-conjugation ($C$), and chiral ($\chi$) symmetries, leaving further generalizations to crystalline TIs for future work. We denote by $c([P])$, the topological index for an equivalence class $[P]$ of adiabatically connected (homotopicaly equivalent) and stably equivalent\footnote{Meaning that $P$'s that differ by the addition of trivial bands are considered equivalent.} spectral projectors $P$, i.e., the projectors are continuously connected while respecting the specified symmetry and the energy gap. The quantity $c([P])$ takes values in an abelian group and is either $\mathbb{Z},\mathbb{Z}_2$ or direct-products thereof if weak-indices are involved\cite{kitaev_periodic_2009,freed2013twisted,stone2010symmetries}. For a composite, but decoupled, 2-component system specified by two spectral projectors $P_1$ and $P_2$ that are of the \emph{same dimensionality} and symmetry, the total topological index satisfies
\begin{align}
c([P_1 \oplus P_2]) = c([P_1]) + c([P_2]).
\label{eqn:sum_c}
\end{align}
Crucially, if $P'$ denotes the total spectral projector after coupling the subsystems,
then $[P_1 \oplus P_2] = [P']$ whenever the coupling maintains the symmetry and the energy gap in the total system.


\paragraph*{Embedded Topological Insulators} If we consider the situation in Fig.~\ref{fig:schematic}(a) where $\text{dim} (\mathcal{M}) > \text{dim} (\mathcal{N})$, then Eqn.~(\ref{eqn:sum_c}) no longer applies or even makes sense.
Nevertheless, we expect that--at least under certain conditions--the system should retain some signature of of the lower-dimensional topological indices associated with $\mathcal{N}.$
For our case of interest, the environment is trivial with
\begin{math}
  c([P_\mathcal{M}]) = 0,
\end{math}
and the embedded subsystem is topological with
\begin{math}
  c([P_\mathcal{N}]) \neq 0,
\end{math}
and assuming no coupling between them, a physical cut that divides the system and cuts through $\mathcal{N}$ will certainly exhibit topologically-protected modes associated with $\mathcal{N}$ as shown in Fig.~\ref{fig:schematic}(b).
Moreover, in this limit the total spectral projector is
\begin{math}
  P = P_\mathcal{M}\oplus P_\mathcal{N}
\end{math}
so that the topological character originates only from $\mathcal{N}$.
The key question then, is what happens when the two systems are coupled? When there is coupling between $\mathcal{M}$ and $\mathcal{N}$,
$P$ is no longer simply the direct sum,
but we naively expect the surface states to survive if the $P_\mathcal{N}$ gap remains open and no symmetries are violated.
We also expect that strong coupling may destabilize the embedded TI $\mathcal{N}$. 
We can make these intuitive statements precise using a disentangling procedure.

\begin{figure*}
\includegraphics[width=0.9\textwidth]{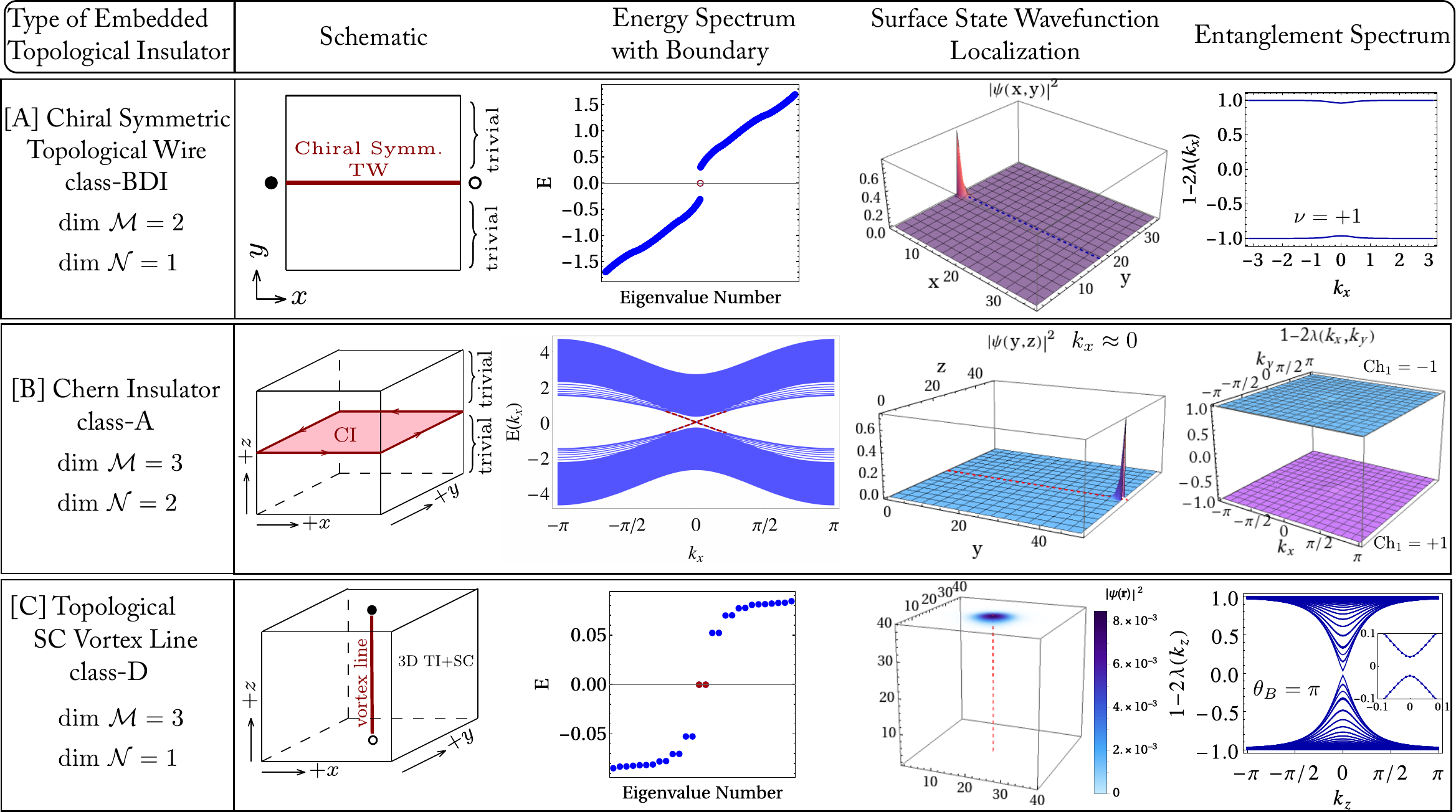}
\caption{(color online) 
Three examples of isolated Embedded Topological Insulators (ETIs) with various symmetry classes and dimensions. Types of ETIs are [A] chiral symmetric topological wire (class-BDI), [B] Chern insulator \cite{bernevig-hughes_book} (class-A), and [C] topological superconducting vortex line \cite{hosur2011majorana} (class-D). Shown are key pieces of information\footnote{Complete details regarding Hamiltonians and parameter values for each example is described in the supplementary material.} regarding these composite insulators organized in tabular form. (Column 1) Schematics for the construction of each composite insulator with the dots and arrowed lines denoting topological surface modes. (Column 2) Energy spectra from open systems with topological modes highlighted as red dots and red lines. (Column 3) Spatial distributions of eigenstate wavefunctions corresponding to topological surface states which are strongly localized at the boundary of the ETIs (dashed lines). (Column 4) Gapped entanglement spectra in the respective ETIs with numerically determined topological indices: chiral winding number [A], Chern number [B] and Wilson loop Berry phase [C] respectively. }
\label{fig:Isolated_ETIs}
\end{figure*}


\paragraph*{Disentangling} To test for an ETI, a region $\mathcal{R}$ that is suspected of harboring an ETI is first identified. To identify if an ETI is present in $\mathcal{R}$
we proceed by imposing an entanglement cut \cite{peschel2003calculation,fidkowski2010entanglement,turner2010entanglement,chang2014symmetry} between $\mathcal{R}$ and the complementary region $\mathcal{R}^c$,
leading to a 1-body reduced density matrix $\rho_\mathcal{R}$ obtained by restriction $\rho_\mathcal{R} = P|_\mathcal{R}$.
The eigenvalue spectrum $\{\Li\}$ of the reduced density matrix
is termed the \emph{entanglement spectrum} (ES),
and is distributed over the unit interval,
\begin{math}
\set{\Li} \subset [0,1].
\end{math}
If the ES is gapped at $\lambda=1/2$,
we may identify the set of eigenstates of $\rho_\mathcal{R}$
with \begin{math}
\Li>1/2
\end{math}
as those with the dominant contributions to correlations within $\mathcal{R}$.    
We can project onto these states with the spectral projector
\begin{align}
  \Pi_\mathcal{R} = \sum_{\Li > 1/2} \ket{\Li} \bra{\Li},
\end{align}
and since the restriction does not spoil the global $\Theta$, $C$, or $\chi$ symmetries, we can compute a topological index $c([\Pi_\mathcal{R}])$
\footnote{A short proof of this is given in the supplemental material.}\cite{fukui15_disen}. 
If
\begin{math}
c([\Pi_\mathcal{R}]) \neq 0
\end{math}
then we say $\mathcal{R}$ is topologically non-trivial.
If $\mathcal{R}$ is \emph{minimal} in the sense that any global-symmetry-preserving cut in $\mathcal{R}$ leads to a gapless entanglement spectrum,
then we call $\mathcal{R}$ an \emph{embedded} TI.\footnote{We specify that the cut be symmetry preserving, because cuts that break the symmetry may lead to spurious (for our purposes) gapped entanglement spectra. For example, taking a 1D Su-Schrieffer-Heeger topological chain and cutting within a unit cell instead of between them will violate the symmetry and lead to a gapped entanglement spectrum even when the chain is topological.} This criterion is based on the fact that all physical and entanglement cuts through STIs
lead to topological surface modes in the energy and entanglement spectra
\cite{turner2010entanglement};
an embedded TI shares the latter property
that all entanglement cuts lead to a gapless ES.

It is instructive to view this prescription in comparison to computing the invariant
\begin{math}
c([P])
\end{math}
of a 1-body Hamiltonian $H$ from its spectral projector $P$.
The reduced density matrix is generated by an entanglement Hamiltonian $H_{\text{ent}}$, \ie
\begin{math}
\rho_\mathcal{R} = (\e^{H_\text{ent}}+1)^{-1}
\end{math}
\cite{peschel2003calculation}.
The spectral gap of $H_\text{ent}$ at 0 is analogous to the spectral gap of $H$ at the Fermi energy where $\Li=1/2$.
It is convenient to consider
\begin{math}
1 - 2\Pi_\mathcal{R}
\end{math}
in place of $\rho_\mathcal{R},$
just as we may have considered the flat-band Hamiltonian
\begin{math}
H_\text{flat} = 1 - 2P
\end{math}
in place of the $H$ itself.
This substitution reduces the entanglement entropy $S_\mathcal{R}$ to zero.
Thus, the entanglement spectrum gap implies the existence of a projector $\Pi_\mathcal{R}$
that allows region $\mathcal{R}$ to be continuously decoupled (disentangled) from the environment
\cite{fukui15_disen}.
In symbols, we mean that after adiabatic deformations the following splitting
\begin{align}
P \rightarrow \Pi_\mathcal{R} \oplus {\Pi_{\mathcal{R}^c}}
\label{eqn:P_split}
\end{align}
is possible whenever the ES is gapped.
Furthermore, if $\mathcal{R}$ is an ETI, then it is topologically non-trivial by itself after decoupling.
A gapless entanglement spectrum represents an obstruction to this decoupling,
and therefore an obstruction to computing the topological index.
\paragraph*{Isolated ETIs}
We present three concrete examples of isolated ETIs
of varying dimensions and symmetry classes
summarized in Fig.~\ref{fig:Isolated_ETIs}. In Fig.~\ref{fig:Isolated_ETIs}(a) we show a 1D BDI-class topological wire embedded in a trivial BDI-class environment, in Fig.~\ref{fig:Isolated_ETIs}(b) we show a 2D A-class Chern insulator embedded in a trivial 3D A-class environment, and in Fig.~\ref{fig:Isolated_ETIs} we show a D-class vortex line embedded in a 3D D-class trivial superconductor environment. In all three cases we show the gapped entanglement spectra, the entanglement topological index, the spectra of the topological bound states, and the localized wavefunctions of the topological bound states. See the Supplemental Material for details of the models and parameters chosen.
These results demonstrate that an isolated ETI
with a gapped, topologically non-trivial entanglement spectrum
is accompanied by topologically protected surface modes.

\paragraph*{Multiple ETIs}We may generalize the preceding argument to a system with multiple embedded TIs.
Let $\mathfrak{R}$ be the disjoint union of ETIs $\mathcal{R}_i$,
\begin{equation}
\mathfrak{R}
= \bigsqcup_{i=1}^N \mathcal{R}_i
= \set{\mathcal{R}_1,\ldots,\mathcal{R}_N};
\end{equation}
then the composite topological index of the system is
\begin{equation}
\label{eqn:cP}
c([P],\mathfrak{R})
= \left(c([\Pi_{\mathcal{R}_1} ]),\ldots,c([\Pi_{\mathcal{R}_N} ])\right),
\end{equation}
where $P$ is the total system spectral projector.
The equivalence classes $[P]$ must now be defined with respect to adiabatic deformations that preserve the ES gaps for each individual cut of the components of $\mathfrak{R}$, just as they require the energy gap to be preserved.
A non-trivial composite index
\begin{math}
c([P],\mathfrak{R}) \neq 0
\end{math}
indicates that the system may be adiabatically decoupled into a collection of lower-dimensional strong TIs.
This index is protected by symmetry and the set of ES and energy gaps.
Eqn.~\eqref{eqn:cP} generalizes the entanglement topological indices of
Refs.~\cite{fukui_entanglement_2014,fukui15_disen,fukui_spin_2016}, and also encompasses other classifications of composite systems
described below. Let us consider some simple examples where the multiple ETIs are arrayed in a translation invariant way.

\paragraph*{Weak TIs}
A WTI can be treated as a crystal of ETIs: the set $\mathfrak{R}$ of ETIs are the layers of the weak TI comprising a lattice.
The composite topological index defined in Eqn.~\eqref{eqn:cP}
is a sequence of indices that repeats after shifting by the number of layers comprising a unit cell, and where
the sum of the indices in each unit cell is non-vanishing.
Because the WTI is necessarily translation invariant,
the composite index can be summarized by a single index for each lattice vector.
\cite{fu_topological_2007}
Any (transverse) cut through the members of $\mathfrak{R}$ leads to surface states,
which accounts for the modes on certain surfaces of the weak TI.
The composite index also accounts for modes bound at certain dislocation defects,
which, at least for edge dislocations, can be treated as additional, truncated ETI subsystems in $\mathfrak{R}$.

\paragraph*{Antiferromagnetic TIs}
An antiferromagnetic TI (AFTI) is another case of an ETI crystal.
The AFTI breaks time-reversal symmetry $\Theta,$
but is symmetric under a magnetic space group
\cite{dresselhaus2007group,mong_antiferromagnetic_2010,vsmejkal2017topological}
generated by the combination
\begin{math}
S = \Theta T_{1/2}
\end{math}
of $\Theta$ and a half-lattice-vector translation $T_{1/2}$,
with $S^2=-1$.
An AFTI can arise when AF order is imposed on a 3D strong TI protected by $\Theta,$
or when stacking layers of 2D TIs with alternating Chern number. Surfaces preserving the magnetic space group symmetry $S$ are gapless,
having zero modes protected by Kramers-like degeneracy
in the plane with zero momentum along $T_{1/2}$.
Like the weak TI, the layers of the AFTI can be treated as ETIs,
and the composite index in Eqn.~\eqref{eqn:cP} is a periodic sequence of alternating indices,
where the sum of indices within the unit cell is zero.

In both of these cases the gapless surface states can be thought of as arising from the individual ETI layers, but the symmetries (translation for WTIs, and $S$ for AFTIs) are crucial for these surface states to remain gapless even when the layers are coupled. In general, even without such symmetries we could still classify both of these systems as a multiple ETI configuration. However, a non-trivial composite index in a system with \emph{multiple} ETIs
does not necessarily imply the existence of gapless topological surface modes
if the modes of different subsystems couple to each other.
Generically the coupling between any pair of subsystems is exponentially suppressed
over distances on the order of correlation lengths. For the ETI crystals the separation is smaller than the lattice constant so the gaplessness will not realistically be preserved. However, if the ETIs are dilute then there may still be clear boundary state signatures. We can make this criterion more precise
using the quantum mutual information measure
\begin{math}
I_2[\mathcal{R}_i;\mathcal{R}_j]
\end{math}
for each pair of ETIs
\cite{wolf2008area}.
In the limit that
\begin{math}
I_2[\mathcal{R}_i;\mathcal{R}_j] \rightarrow 0,
\end{math}
the ETIs $\mathcal{R}_i$ and $\mathcal{R}_j$ are independent
with decoupled topological surface modes.
\footnote{
	A simple example of the mutual information
	between two embedded Chern insulators
	is given in the Supplementary Material.
}



\begin{figure}
\includegraphics[width=7.5cm]{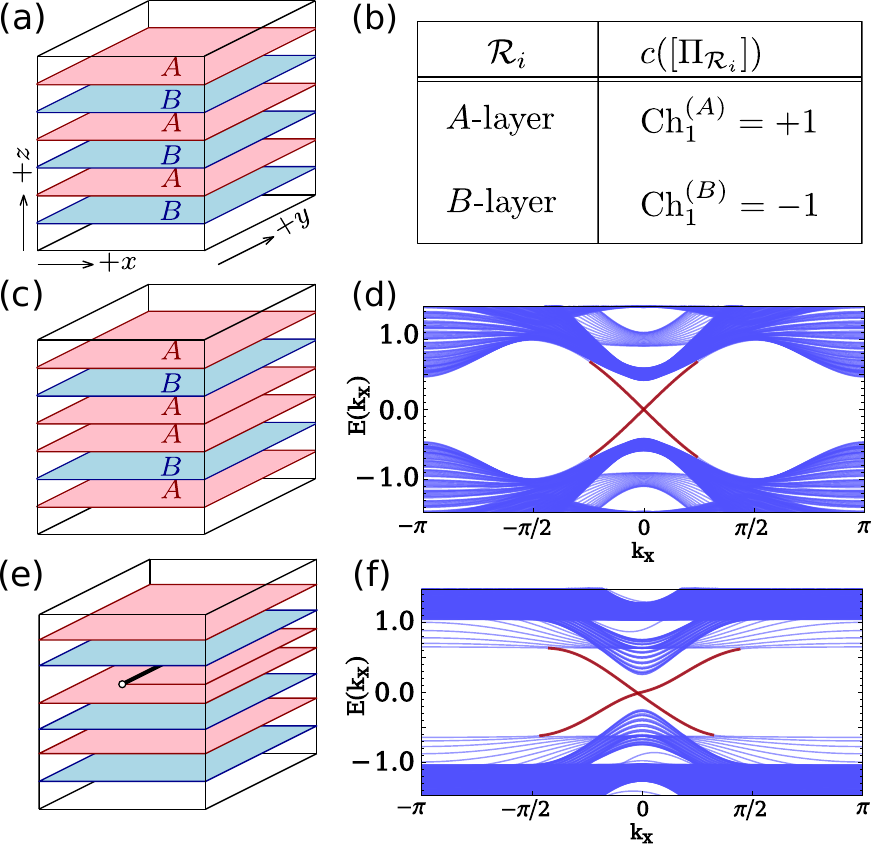}
\caption{\label{fig:stacking}(color online)
  (a) Schematic of the stacked Chern insulator construction which consists of two layers (labeled $A$ and $B$ type) with opposite Chern numbers $\text{Ch}_1=\pm 1$. (b) Table of ETI Chern indices. Diagrams of a (c) stacking fault and a (e) partial dislocation and their corresponding energy spectra (d,f) with the (red lines) topological modes. In (d) there is an open boundary where the topological mode localizes on the defect region. The Hamiltonian and its parameters are described in the supplemental material.}
\end{figure}

\paragraph*{Defects as ETIs}
Despite the instability of surface states generated in generic multiple ETI configurations without additional symmetries, the ETI concept makes new, robust predictions for topological defects. For example, in the vortex line model in Fig.~\ref{fig:Isolated_ETIs}[c]
an ETI naturally arises on a topological defect of the model. Another place where there is an important connection between defects and ETIs is in crystals where the unit cell itself is not topological, but subunits of the cell are topological. Then natural crystal defects, such as stacking faults or partial dislocations, that can break the unit cell structure can leave behind ETI remnants and their associated topological boundary states.

One of the simplest realizations of the latter
is a three-dimensional crystal built from a stack of alternating Chern insulators described in Fig.~\ref{fig:stacking}(a).
The insulator layers have opposite Chern numbers
\begin{math}
\text{Ch}_1 = \pm 1,
\end{math}
and the interlayer hopping amplitudes are chosen to break inversion symmetry, and the AFTI space group symmetry,
so that the crystal is neither a strong TI, nor a weak TI, nor an AFTI,
nor a mirror-symmetric crystalline TI,
yet the crystal has a non-trivial classification by the composite index in Eqn.~\eqref{eqn:cP}. To find a consequence of the non-trivial topology we can consider defects.
The crystal supports lattice defects
including stacking faults, as in Fig.~\ref{fig:stacking}(c),
and partial dislocations, as in Fig.~\ref{fig:stacking}(e).
In both cases,
and although the crystal is topologically trivial in the conventional sense,
the defect is an ETI with localized, topologically-protected modes:
the stacking fault yields modes localized at an open boundary,
as in Fig.~\ref{fig:stacking}(d),
and the dislocation yields modes localized at the defect,
as in Fig.~\ref{fig:stacking}(f).
Such a crystal may be realized in layered antiferromagnetic (AFM) materials
where the AFM order is commensurate to the layer stacking,
The dimerization coupling spoils the magnetic space group symmetry required for an AFTI so the surface states are generically gapped.
However, an AFM domain wall forming a stacking fault can generate gapless chiral modes.
Honeycomb materials such as \mbox{Sn-X~(tin-halides)}~\cite{niu2017quantum}
are promising candidates to realize this proposal
because topological states are predicted to coexist with antiferromagnetic order.
Predicted AFTIs like \mbox{GdPtBi} may also realize ETI crystals
if the magnetic group symmetry is broken
such that topological edge modes only appear on certain domain walls, and not on full surfaces. This provides a new avenue in the search for materials yielding topological phenomena as there will be systems where the surfaces are gapped but certain crystal defects still yield topologically protected modes. In some aspects this is reminiscent, but not identical to the concept of higher-order topological insulators\cite{benalcazar2017a,benalcazar2017b,schindler2017,song2017,langbehn2017}

\paragraph*{Disorder}
We have limited the discussion here to clean ETIs,
but Eqn.~\eqref{eqn:cP} remains valid even when disorder is present.
In this case,
the indices $c([\Pi_{\mathcal{R}_i}])$ must be computed using either
the machinery of non-commutative geometry~\cite{prodan2010non,prodan2016bulk}
or almost commuting operators~\cite{hastings2011topological,loring2015k}
which have been developed for disordered topological insulators
\footnote{We should mention that these generalized method to compute topological invariants have so far only been rigorously proven for a subset of the possible 10-symmetry classes in dimensions 1 to 3 and it remains an on-going effort to fill in the gaps.}.


In conclusion, we have shown that a system with a collection of ETIs is characterized by a composite topological index consisting of the direct product of indices of disjoint subsystems.
Each subsystem retains its topological character--and may be regarded as decoupled--if its mutual information with each of the other subsystems vanishes.
In a more practical setting -- say in the study of complicated heterostructures -- the application of Eqn.~(\ref{eqn:cP}) to ab-initio accurate Density Functional Theory~(DFT) data could be hindered by the need to determine ETI regions which are unknown a-priori. To this end, we propose modifying and utilizing hierarchical clustering algorithms\cite{erciyes2014complex,de2016spectral} from complex networks theory. Specifically, the splitting operation of Eqn.~(\ref{eqn:P_split}) makes this ideal for a divisive type algorithm\cite{kaufman2009finding} whose purpose is to cluster DFT orbitals into different component ETIs and a trivial background.


\paragraph*{Acknowledgements} This work was supported in part by the Gordon and Betty Moore Foundation EPiQS Initiative through Grant GBMF4305 (VC) at the University of Illinois. TLH was supported by the ONR YIP Award N00014-
15-1-2383. We gratefully acknowledge useful discussions with V. Dwivedi.

\bibliography{EmbeddedTI}

%
%
%
%
%
%
%


\onecolumngrid

\section{Supplementary Material for ``Embedded Topological Insulators''}
The supplementary materials collected here contain technical details that were omitted from the main text. In Sec.~I we provide a short proof that restriction or partial tracing states in real-space preserves global symmetries. In Sec.~II we describe in detail the various isolated Embedded Topological Insulator (ETI) models appearing in Fig.~2 of the main text. In Sec.~III we provide further details regarding the situation of two ETIs of opposite topological indices placed in proximity to each other and quantify their interaction through the quantum mutual information measure. Next in Sec.~IV we present detailed model information regarding the ETI crystal and its defects as discussed in the main text. Then in Sec.~V we give further introductory details regarding the conventional classification of topological insulators using spectral projectors onto occupied states. Lastly, in Sec.~VI we supply a short supplement regarding the intricacies of computing the winding number topological index for chiral symmetric wires.   
\\\\
\twocolumngrid

\section{I. Restriction Preserves Global Symmetries}
\label{supp:restrict}

Implicitly assumed in the main text is the fact that the entanglement isolated spectral projector $\Pi_\mathcal{R}$ maintains the global symmetries of the system. In this supplementary section we provide short a proof of this fact.
The global symmetry implemented by the linear operator $g$ acts on the total Hamiltonian as $[H,g]_{\pm}=0$. Whether or not the commutator $[\cdot,\cdot]_+$ or anti-commutator $[\cdot,\cdot]_-$ is used depends on the relevant symmetry, which is either commuting or anti-commuting. The later case is relevant to charge-conjugation and chiral-sublattice symmetries. The occupied and unoccupied states are respectively degenerate under the flat band Hamiltonian $H'=\mathbbm{1}-2P$ with eigenvalues $-1$ (occupied) or $+1$ (unoccupied).
The flat band Hamiltonian also has the symmetry $[\mathbbm{1}-2P,g]_\pm=0$ because
a commuting symmetry maps each state to another with the same eigenvalue under $H$ or $H'$
while an anticommuting symmetry maps each state to one with the opposite eigenvalue under $H$ or $H'$.
A real-space bipartition into $\mathcal{R}$ and its complement $\mathcal{S}=\mathcal{R}^c$ splits the total Hilbert space into $\mathcal{H}= \mathcal{H}_\mathcal{R}\oplus\mathcal{H}_\mathcal{S}$. We can then divide the flat band Hamiltonian $H'$ into three contributions
\begin{align}
H'= H'_\mathcal{R} + H'_{S} + H'_{\mathcal{RS}}
\end{align}
where the first two terms have support in $\mathcal{H}_\mathcal{R}$ and $\mathcal{H}_\mathcal{S}$ respectively. The last term represents the coupling between $\mathcal{R}$ and $\mathcal{S}$.
Because $g$ is a global (on-site) symmetry, it cannot change the support of each term in $H'$; therefore, it is not only a symmetry of $H'$, but also a symmetry of each term individually, \ie
\begin{equation*}
  [H'_\mathcal{R},g]_\pm = [H'_\mathcal{S},g]_\pm = [H'_\mathcal{RS},g]_\pm = 0.
\end{equation*}
The flat band Hamiltonian restricted to $\mathcal{R}$ is given through the restricted single-particle correlator $\rho_\mathcal{R}$,
\begin{equation*}
  H'_\mathcal{R} = \mathbbm{1}_\mathcal{R}-2\rho_\mathcal{R}.
\end{equation*}
By the same argument that the Hamiltonian $H$ its flattened counterpart $H'$ have the same symmetry,
the flattened counterpart of $H'_\mathcal{R}$ will have the same symmetry, or
\begin{equation*}
  [g,\mathbbm{1}_\mathcal{R}-2\Pi_\mathcal{R}]_\pm=0
\end{equation*}
where $\Pi_\mathcal{R}$ is the flatten form of $\rho_\mathcal{R}$.
This proves our assertion that restriction and subsequent flattening preserves global symmetries.


\section{II. Isolated Embedded Topological Insulators: Models and Numerics}
\label{supp:All_Isolated_ETIs}

\subsection{IIA. Embedded Chiral Symmetric Wires in 2D}
\label{supp:ETI_CW}



\begin{figure}
\boxed{\includegraphics[width=0.28\textwidth]{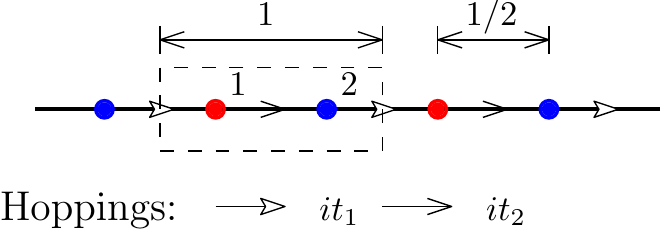}}
\caption{(color online) A two orbital model of 1D wire with chiral symmetry from the BDI class. The dots 1 (red) and 2 (blue) label the different orbital states. The dashed box denotes the unit-cell and the arrows the direction of the imaginary valued nearest neighbor hopping matrix elements. We take the unit-cell to be of length 1.}
\label{fig:BDI_wire}
\end{figure}

In this section we discuss one of the simplest Embedded Topological Insulator (ETI) which consists of a topological chiral symmetric wire embedded in a trivial 2D environment. This symmetry class is commonly associated with a bipartite lattice and Hamiltonians that respect this symmetry have no matrix elements between orbitals of the same sublattice type. More precisely, if $H$ is the 1-body Hamiltonian and $S$ is the chiral charge operator that is $+\mathbbm{1}$ on one sublattice type but $-\mathbbm{1}$ on the other, then $H$ has chiral symmetry whenever $\{H,S\}=0$. We remark that in one dimension the chiral classes AIII, BDI and CII are special in that they are the only symmetry classes that support strong topological insulators that are outside of the superconducting Bogoliubov-deGennes (BdG) ensembles. In particular, the 1D AIII and BDI classes carry $\mathbb{Z}$ valued invariants while CII has $2\mathbb{Z}$ valued invariants from Kramers degeneracy.

We begin with a minimal 1D wire model with BDI chiral symmetry as defined in Fig.~\ref{fig:BDI_wire}. We have chosen the BDI class for reasons of simplicity. Time-reversal symmetry is realized as $T^2=1$ and charge-conjugation by complex conjugation $P=K$. Written in a two-component basis, the Bloch Hamiltonian takes the form
\begin{align}
H_\text{\tiny1D}(k_x) = (t_1+t_2) \sigma^x \sin (\tfrac{k_x}{2}) - (t_1 - t_2) \sigma^y\cos (\tfrac{k_x}{2})
\end{align}
in the gauge with quasi-periodic Bloch eigenstates $|u_n(k_x + 2\pi)\rangle = \e^{- i 2\pi \hat{r}_x} |u_n(k_x)\rangle$ where $\hat{r}_x=\tfrac{1}{2}\sigma^z$ is the orbital position operator within the unit-cell. The Hamiltonian maintains chiral symmetry with $S=\sigma^z$ and time-reversal symmetry with $T = K \sigma^z$. The strong index $\nu \in \mathbb{Z}$ which counts the number of zero energy modes at an open boundary is~\cite{ryu_topological_2010,mondragon-shem_topological_2014}
\begin{align}
\nu = \int \frac{\diff k_x}{\pi}\sum_{n}
\langle u_{n}(k_x)| i\,S\,\nabla_{k_x} u_n(k_x)\rangle
\end{align}
where the sum extends over occupied Bloch bands and the covariant derivative is $\nabla_{k_x} = \partial_{k_x} + i \hat{r}_x$. In practice, $\nu$ is more easily determined from its property as a winding number. More details of this and other subtleties regarding gauge and unit-cell choices are discussed in the Supplementary Section VI.
In this model the parameter range $|t_1| > |t_2|$ leads to a strong topological phase with $\nu = 1$ and otherwise a trivial one when $|t_1| < |t_2|$. We will term the strong topological phase a topological chiral symmetric wire.

\begin{figure}
\boxed{\includegraphics[width=0.3\textwidth]{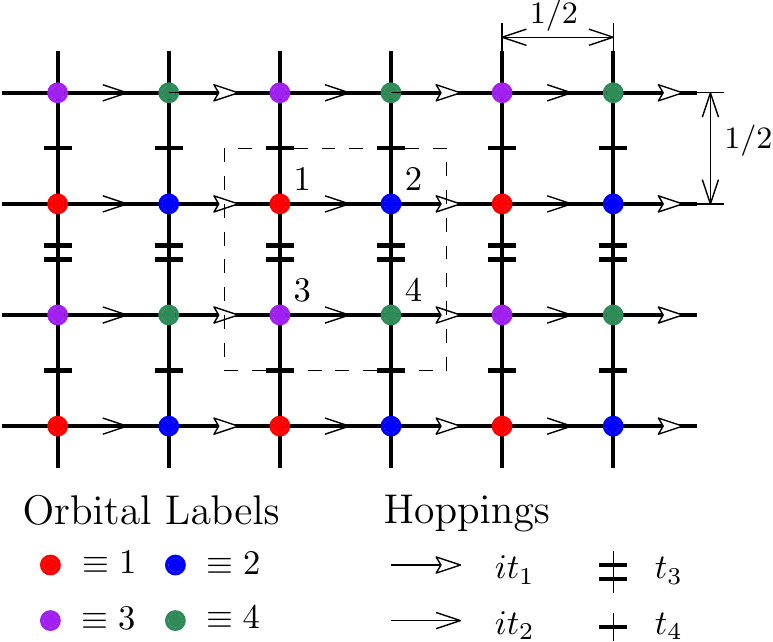}}
\caption{(color online) The square lattice system constructed from stacking 1D chiral symmetric wires. The dashed square denotes the boundary of the unit-cell and there are 4 orbitals in a unit-cell labeled by the different colored dots. The real parameters $t_3,t_4$ are the intra unit-cell and inter unit-cell vertical hoppings respectively.}
\label{fig:BDI_stacked_wires}
\end{figure}

A 2D model with chiral symmetry is constructed by stacking 1D wires into a 2D square lattice as is shown in Fig.~\ref{fig:BDI_stacked_wires}. Because of the requirements of chiral symmetry, the unit-cell has to be enlarged to contain 4 orbitals with two 1D `sub-wires'. The chiral charge operator $S$ then alternates its sign between sub-wires within the unit-cell, \ie $S=\sigma^z \otimes \sigma^z$. The Bloch Hamiltonian for this model is
\begin{align}
\label{eqn:CI_Ham}
H_\text{\tiny2D}(\kk) &=
t^+_{12} (\mathbbm{1}\otimes \sigma^x) \sin (\tfrac{k_x}{2})
-t^-_{12} (\mathbbm{1}\otimes \sigma^y)\cos (\tfrac{k_x}{2})
\nonumber \\
&+t^+_{34} (\sigma^x\otimes\mathbbm{1}) \cos(\tfrac{k_y}{2})
+t^-_{34} (\sigma^y\otimes \mathbbm{1}) \sin(\tfrac{k_y}{2})
\end{align}
where $t^\pm_{ij} = t_i \pm t_j$. We make this model \emph{trivial} by trivializing each wire: limiting to parameter ranges such that $|t_1|<|t_2|$ and $|t_3|,|t_4|< \text{min}(|t_1|,|t_2|)$.

\begin{figure}
\includegraphics[width=3.5cm]{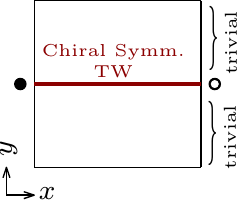}
\caption{(color online) Schematic of an embedded chiral topological wire. The dark red line denotes the location of an `impurity layer' containing a single topological wire of the type in Fig.~\ref{fig:BDI_wire} with parameters $|t'_1| > |t'_2|$ surrounded by a trivial insulator model of the type in Fig.~\ref{fig:BDI_stacked_wires} with $|t_1|<|t_2|$ and $|t_4|<|t_3|$. The dots denote zero modes that are expected to be present and localized the at open boundaries.}
\label{fig:ETI_CW}
\end{figure}

\begin{figure}
\includegraphics[width=8cm]{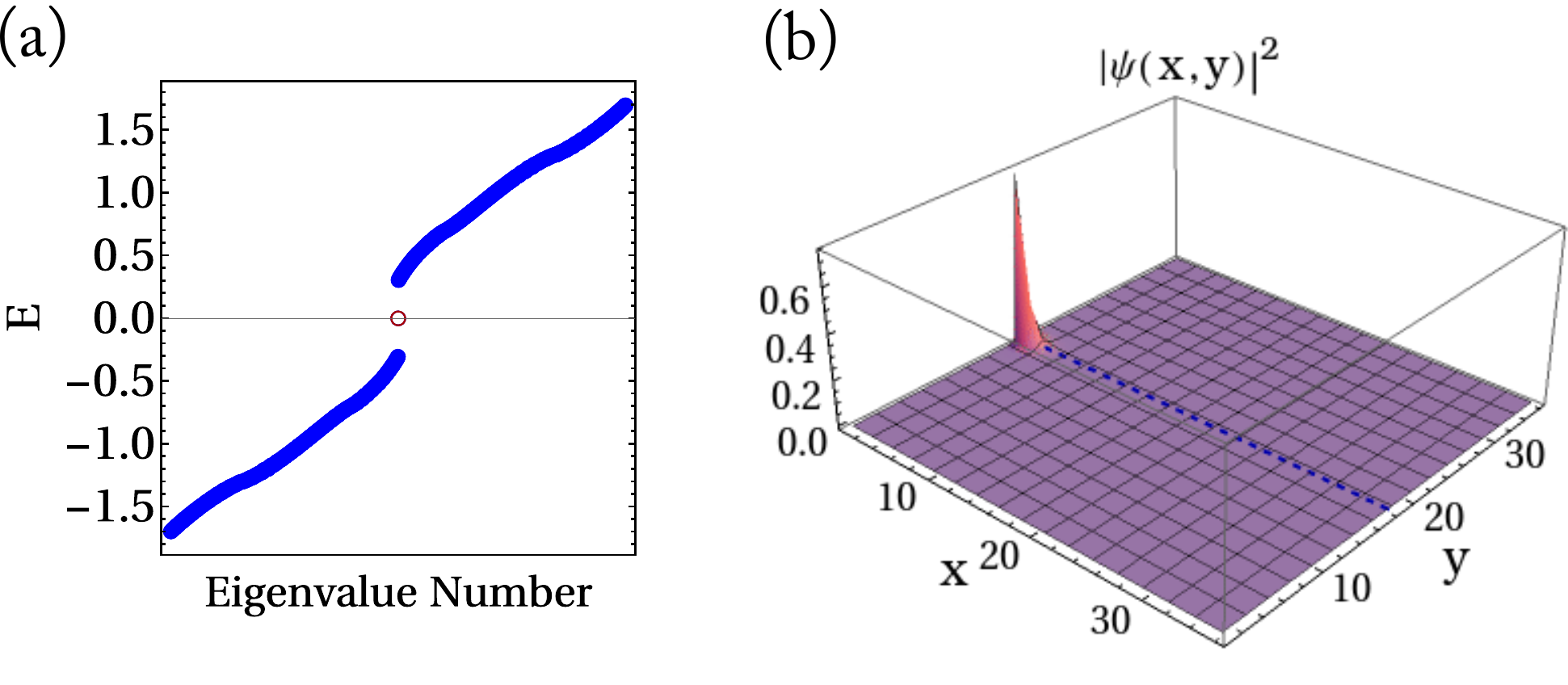}
\caption{(color online) Exact diagonalization results for a square system with linear dimensions $L_x=L_y=36$ and containing a single embedded topological chiral wire and open boundaries along $x$. Parameters are $(t_1,t_2,t_3,t_4) = (0.5,1.0,0.1,0.1)$ for the trivial square lattice environment and $(t_1',t_2') = (1.0,0.5)$ for the single impurity wire that is the embedded chiral topological wire. (a) Energy spectrum with two zero energy modes (open red circles). (b) Wavefunction amplitude distribution for a zero mode. The dashed (dark blue) line denotes the location of the impurity layer.}
\label{fig:energy_wfunc}
\end{figure}

\begin{figure}
\includegraphics[width=4cm]{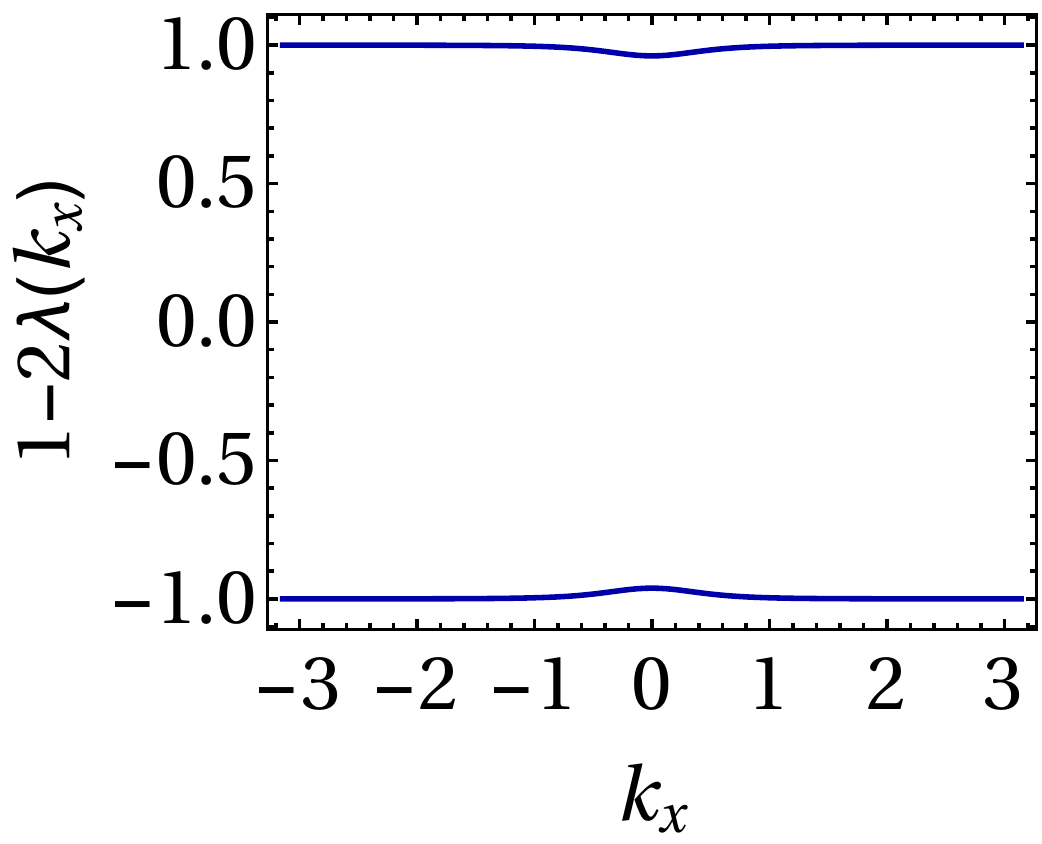}
\caption{(color online) The entanglement spectrum obtained by restricting to just the single embedded topological chiral wire. Computing the strong invariant $\nu$ from these entanglement spectrum bands yields $\nu=1$ in agreement with the appearance of physical zero modes in the inhomogeneous system.}
\label{fig:ETI_ECW_ES}
\end{figure}

To construct an embedded topological chiral wire system, we replace a single trivial sub-wire within one unit-cell with hopping parameters $t_1',t_2'$ such that $t_1'>t_2'$. Schematically the system we have in mind is shown in Fig.~\ref{fig:ETI_CW}. Numerical exact diagonalization of this inhomogeneous system leads to the energy spectrum in Fig.~\ref{fig:energy_wfunc}(a) which clearly shows the presence of localized zero energy modes as demonstrated in Fig.~\ref{fig:energy_wfunc}(b).
Finally, the entanglement spectrum of a closed system (Fig.~\ref{fig:ETI_ECW_ES}) is gapped, with the expected non-trivial band topology with $\nu=1$ derived from the entanglement band wavefunctions.


\subsection{IIB. Embedded Chern Insulator in 3D}
\label{supp:ETI_CI}


\begin{figure}[ht] 
\includegraphics[width=4.4cm]{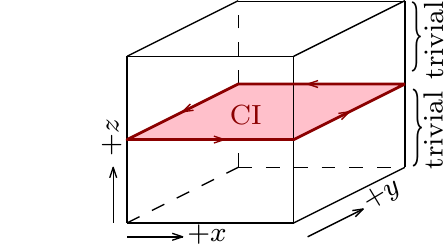}
\caption{(color online) Schematic of the embedded Chern Insulator (ECI). The shaded (red) `impurity' layer is the ECI sandwiched by trivial layers. Arrows denote the direction of the chiral edge currents for open boundaries in the $x,y$ directions.}
\label{fig:ETI_CI}
\end{figure}

\begin{figure}
\includegraphics[width=4.2cm]{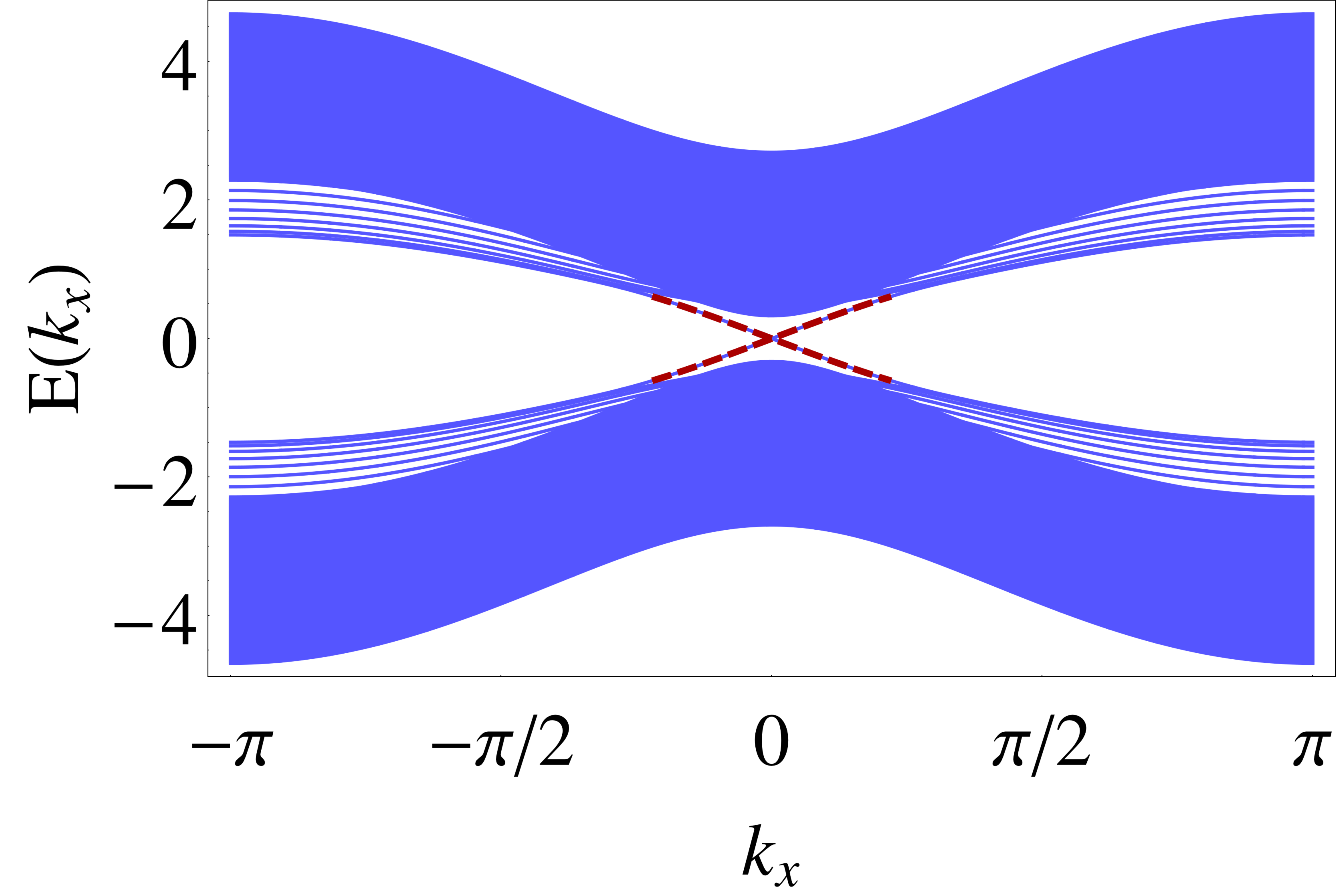}
\caption{(color online) Energy band structure of an ECI crystal infinite in $x$ but with finite dimensions $L_y=L_z=20$ and has PBC in $z$, OBC in $y$. Parameters are $(m,m',\delta_z)=(-0.5,0.5,0.1)$. Thick bands (blue) are the 3D bulk bands, thin lines (blue) are 2D-like that localize near the CI layer. Dashed lines (red) are the topological edge modes.}
\label{fig:ETI_CI_BS}
\end{figure}

In this section we describe in further detail the Embedded Topological Insulator (ETI) formed by a 2D Chern Insulator (CI) embedded in a trivial 3D environment.
This is yet another example of a co-dimension-1 ETI, but in 3D.
The 2D CI falls under Altland-Zirnbauer class-$A$ and has a $\mathbb{Z}$ valued invariant.
A schematic of such an inhomogeneous crystal is shown in Fig.~\ref{fig:ETI_CI}; the CI is marked as an ``impurity layer'' in an otherwise-pristine crystal formed by stacking two-dimensional layers.
In the pristine 3D environment the two orbital model Bloch Hamiltonian is taken to be\cite{bernevig-hughes_book}
\begin{align}
H(\kk) = &\sin k_x \sigma^x + \sin k_y \sigma^y
+ \left(2-m-\cos k_x -\cos k_y  \right)\sigma^z \nonumber \\
&+ 2 \delta_z \cos k_z\, \mathbbm{I}_2
\end{align}
where $\delta_z$, $m$ are real parameters. The coupling between $xy$-layers in the $+z$ direction is quantified by $\delta_z$. When $\delta_z=0$, each 2D $xy$-layer is a representative model of a CI whenever $0<m<2$ (Ch$_1=+1$) or $2<m<4$ (Ch$_1=-1$) and is otherwise trivial (Ch$_1=0$) when $m<0$ or $m>4$. We set $m<0$ for the trivial bulk but let $0 < m'<2$ in the CI layer (See Fig.~\ref{fig:ETI_CI}).

\begin{figure}
\includegraphics[width=4.0cm]{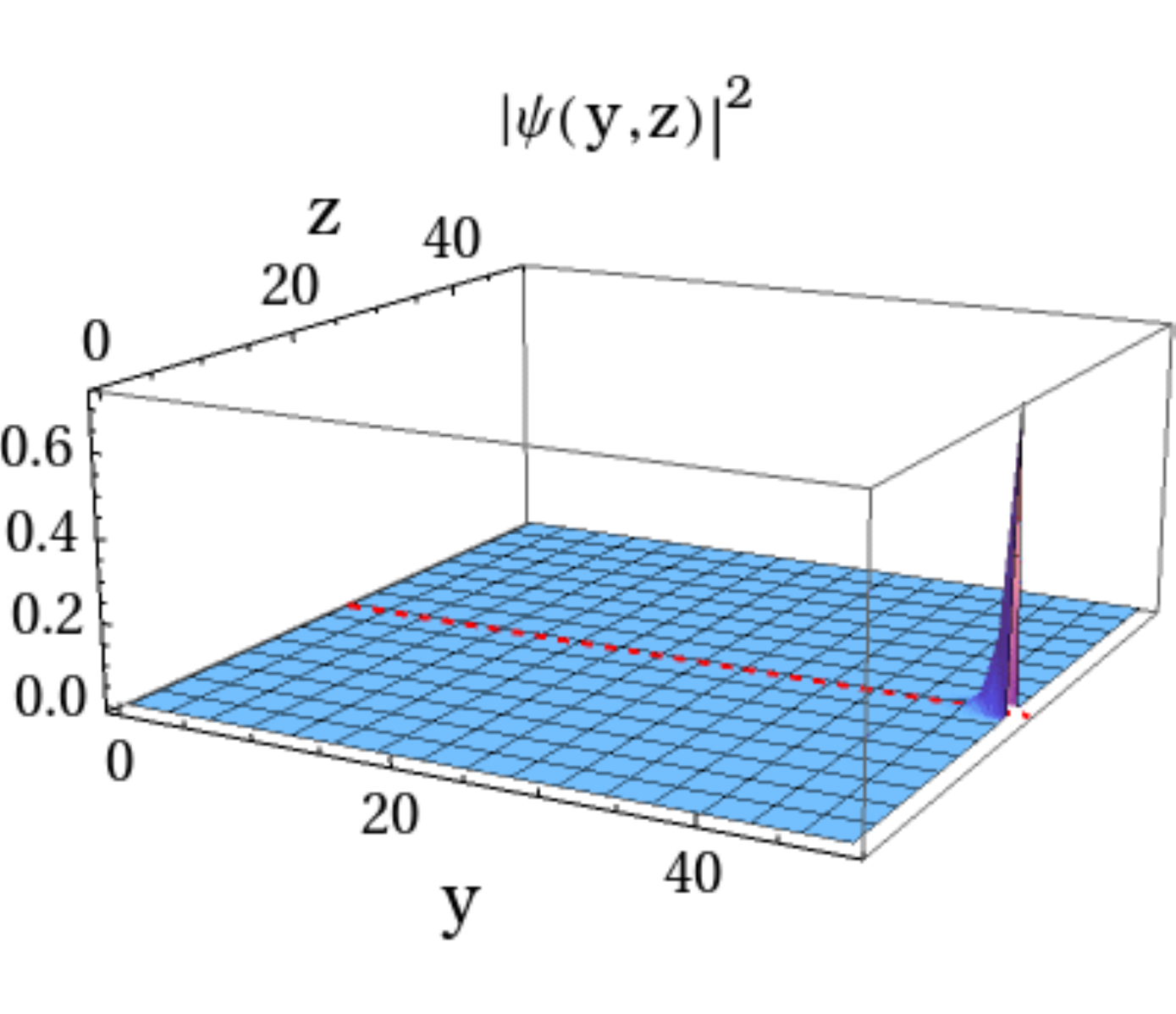}
\caption{(color online) The wavefunction distribution of a localized chiral edge mode. Dashed (red) line denotes the location of the CI layer and $L_y=L_z=48$ with fixed $k_x\approx =0$.}
\label{fig:ETI_CI_PSI2}
\end{figure}

For small values of $\delta_z$, typically $|\delta_z| < |m|,|m'|$, the total system retains its energy gap between conduction and valence bands, while remaining adiabatically connected to the decoupled limit $\delta_z=0$. This precise qualification of adiabatic continuity is revealed through the disentangling method that we have implemented to identify the embedded Chern Insulator (ECI).
Figure~\ref{fig:ETI_CI_BS} shows an open boundary energy band structure of the ECI with the tell-tale chiral edge modes, confirming that the total system is topologically non-trivial. The wavefunction weight $|\psi|^2$ of a chiral edge mode is shown in Fig.~\ref{fig:ETI_CI_PSI2} and verifies the localization on the edge near the CI layer.

\begin{figure}
\includegraphics[width=3.8cm]{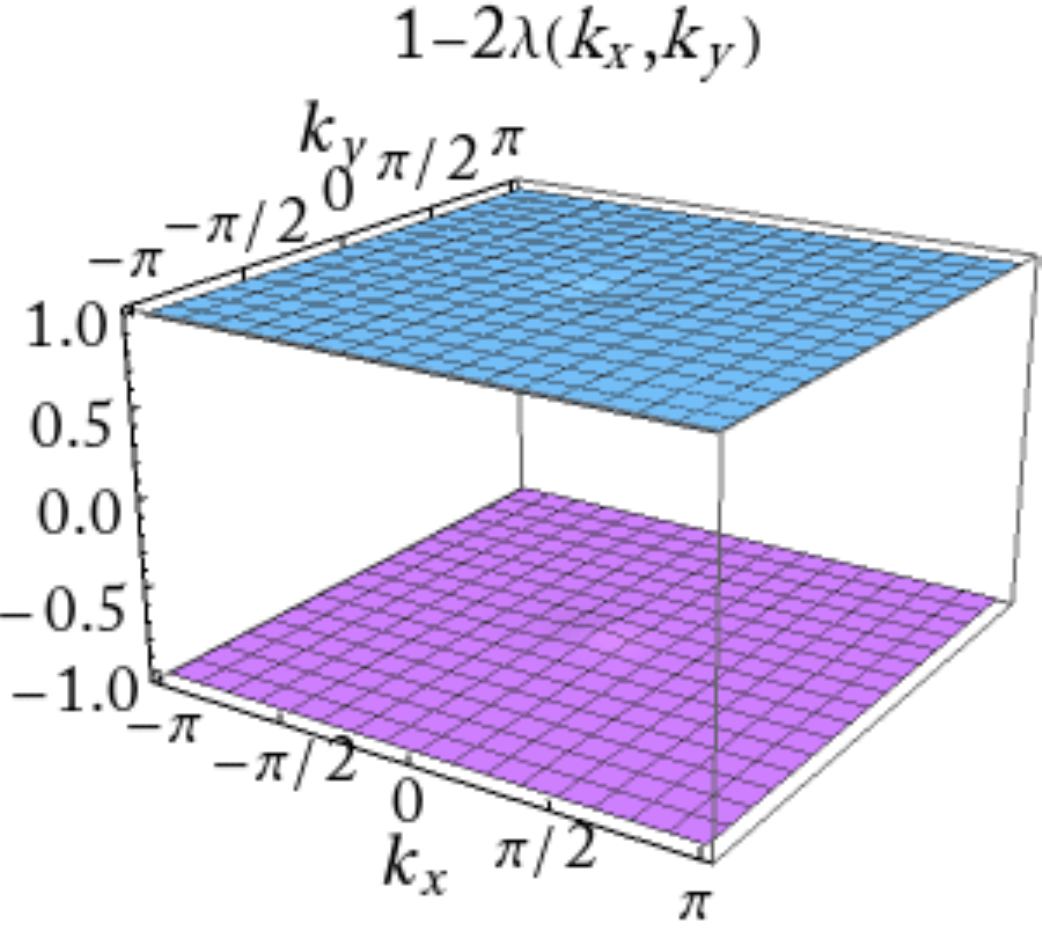}
\caption{(color online) The gapped entanglement spectrum band structure of the isolated CI layer with $L_z=20$ and infinite in $x,y$. Numerical calculation of the TKNN number from just the lower occupied band gives Ch$_1=1$. 
}
\label{fig:ETI_CI_ES}
\end{figure}

Tracing out all $xy$-layers save the one containing the ECI one produces the gapped entanglement band structure shown in Fig.~\ref{fig:ETI_CI_ES} that is topologically non-trivial with Ch$_1=+1(-1)$ in its lower (upper) band. We note that tracing over other individual layers or collection of layers produces a similarly gapped entanglement band structure but with zero Ch$_1$ unless the CI layer is included.

\section{IIC. Embedded Topological Vortex in 3D}
\label{supp:ETI_V}


\begin{figure}
\includegraphics[width=4.4cm]{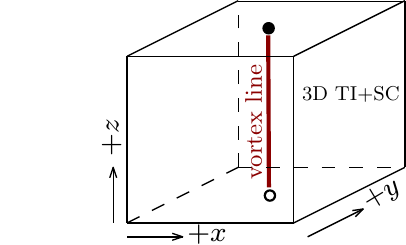}
\caption{(color online) Schematic of an embedded topological vortex within a trivial superconductor with spin-orbit coupling. The bulk superconductor is formed by introducing s-wave pairing to a time-reversal invariant strong topological insulator. The (red) line is the vortex within the trivial superconductor.}
\label{fig:ETI_V}
\end{figure}

\begin{figure}
\includegraphics[width=6.5cm]{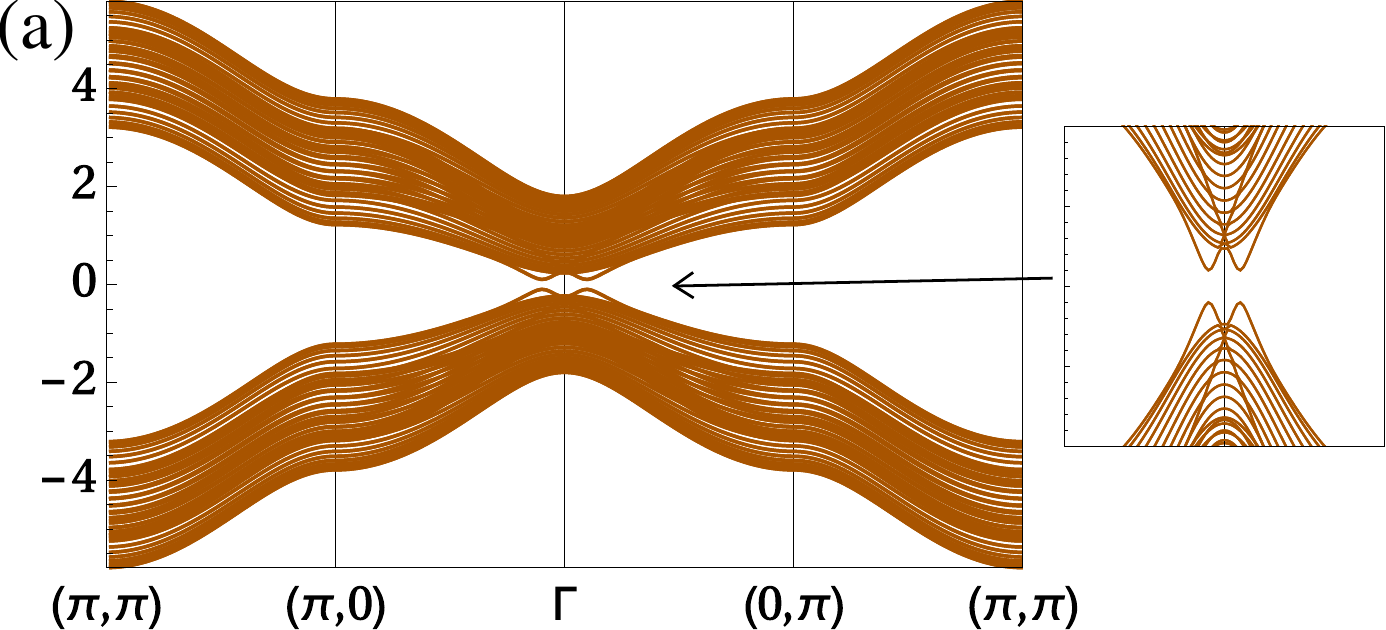}
\caption{(color online) The [001] surface energy BdG band structure of a trivial superconductor formed from a strong time-reversal invariant topological insulator with uniform s-wave superconducting pairing. The model parameters values are $(t,m_0,M,E_F,\Delta)=(0.5,-1.0,2.5,0.3,0.1)$. Inset shows that the topologically protected surface Dirac codes at the $\Gamma$ point are gapped out by the superconducting pair potential.} 
\label{fig:ETI_V_BS}
\end{figure}


In this section we describe the model of the embedded topological vortex (ETV) that arises as a topological defect in a three dimensional s-wave superconductor (SC) with spin-orbit coupling (Fig.~\ref{fig:ETI_V}). Specifically, the model is the one introduced in Ref.~\onlinecite{hosur2011majorana} and can be best described as a time-reversal invariant strong topological insulator with s-wave pairing. The mean-field Bogoliubov-de Gennes (BdG) Hamiltonian has the following form
\begin{align}
H=\frac{1}{2} \sum_\mathbf{k} \Psi^\dagger_\mathbf{k} \mathcal{H}_\text{BdG}(\mathbf{k}) \Psi_\mathbf{k}
\end{align}
with 
\begin{subequations}
\begin{align}
&\Psi^\dagger_\mathbf{k} =
(c^\dagger_{\alpha\uparrow\mathbf{k}}, \;
c^\dagger_{\alpha\downarrow\mathbf{k}}, \;
c_{\alpha\downarrow\mathbf{-k}},\;
-c_{\alpha\uparrow\mathbf{-k}}), \quad \alpha =\uparrow,\downarrow\\
&\mathcal{H}_\text{BdG}(\mathbf{k}) =
\begin{pmatrix}
H_\text{STI}(\mathbf{k})-E_F & \Delta \\
\Delta^* & E_F- H_\text{STI}(\mathbf{k})
\end{pmatrix},
\end{align}
\end{subequations}
where $E_F$ is the chemical potential and $H_\text{STI}(\mathbf{k})$ is the Bloch Hamiltonian of the strong time-reversal invariant topological insulator. $H_\text{STI}$ describes a 2-orbital tight-binding model on a cubic lattice with spin-1/2 electrons and is given by
\begin{subequations}
\begin{align}
H_\text{STI}(\mathbf{k}) &=
(\vec{d}(\mathbf{k})\cdot \vec{\sigma})\otimes \tau^x +  m(\mathbf{k})\,\mathbbm{1}\otimes \tau^z \\
d_i (\mathbf{k}) &= 2t \, \sin k_i \, \sigma^i, \quad i = x,y,z \\
m(\mathbf{k}) &= M + m_0 \sum_{i=x,y,z} \cos k_i
\end{align}
\end{subequations}
with $t,M,m_0$ being real valued parameters. The Pauli matrices $\tau^\alpha$ act on the orbital degrees of freedom while $\sigma^i$ acts on spin. The non-superconducting strong  topological insulator (STI) phase corresponds to the parameter range $-3 < \frac{M}{m_0}< -1$ and is due to spin-orbit coupling. Time-reversal symmetry is manifested by the condition
\begin{align}
(\sigma^y\otimes \mathbbm{1})\, {H_\text{STI}(-\mathbf{k})}^* \,(\sigma^y\otimes \mathbbm{1}) = H_\text{STI}(\mathbf{k}).
\end{align}

\begin{figure}
\includegraphics[width=8cm]{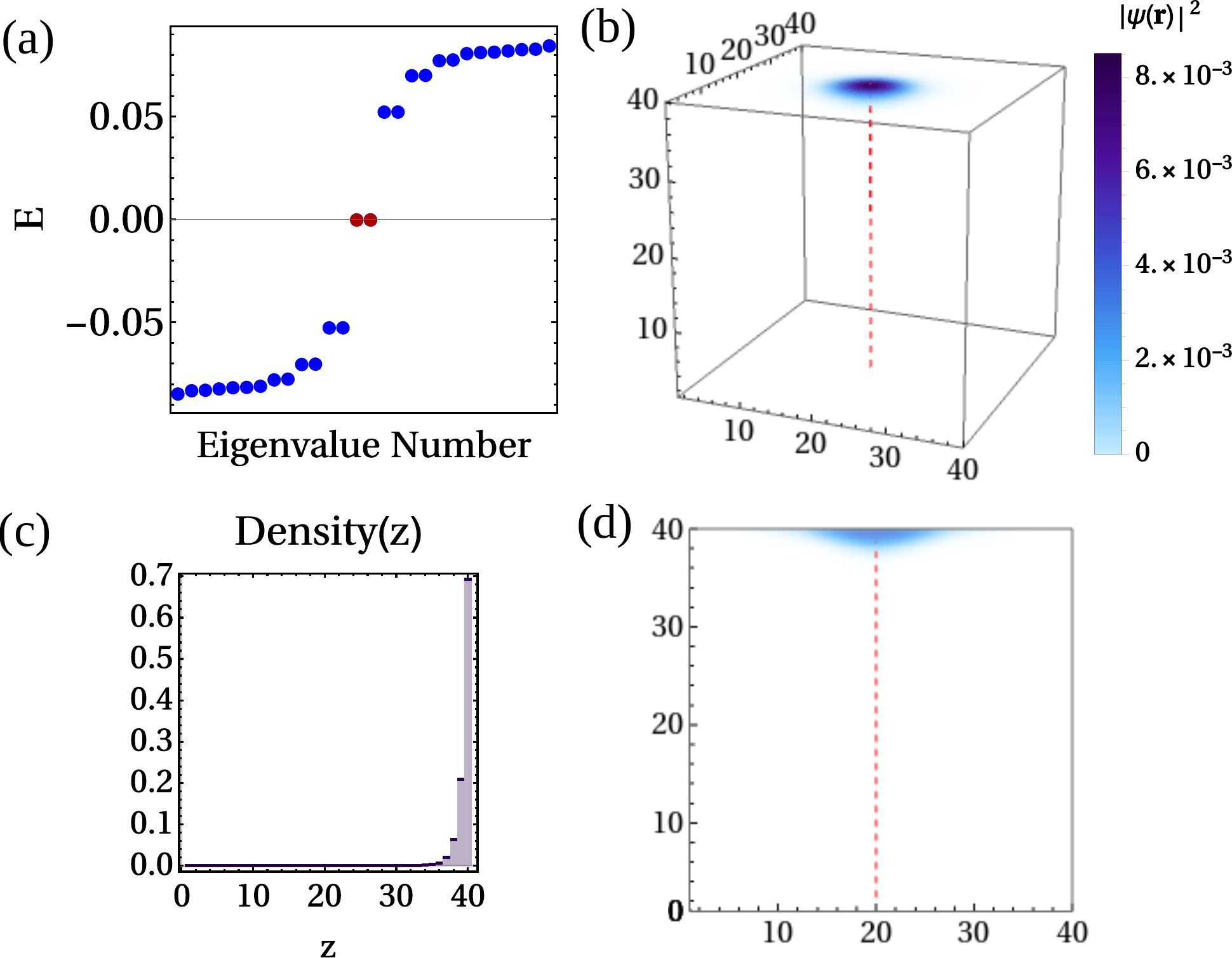}
\caption{(color online) Majorana zero modes (MZMs) bound to the ends of a vortex in the trivial D-class superconductor with spin-orbit coupling. Numerical results are from exact diagonalization of a $40\times 40 \times 40$ crystal with open boundaries in $x,y,z$ and a vortex line (dashed red line) placed at the center of the $x,y$ plane according to Eqn. (\ref{eqn:vortex_delta}). Model parameter values are the same as in Fig.~\ref{fig:ETI_V_BS} with $\Delta_0 = 0.1$ and $\xi = 2$. (a) Energy spectrum of the total system near zero frequency. Red dots denote the pair of MZMs localized on the top and bottom surfaces. (b) Wavefunction distribution of a MZM strongly localized on the top surface with a side view in (d) and integrated layer density in (c).}
\label{fig:MZM}
\end{figure}

In the pristine case, the superconducting order parameter $\Delta$ is taken to be a constant. Due to the existence of superconducting pair correlations, the normally present topologically protected surface Dirac cones are gapped out as is shown in Fig.~\ref{fig:ETI_V_BS} whenever $E_F$ lies in the normal phase band gap. Moreover, in the absence of vortices the surface energy bands remain gapped even when the normal phase becomes metallic whenever $E_F$ lies in the conduction or valence bands. Thus the bulk 3D SC as described by $\mathcal{H}_\text{BdG}$ is a trivial superconductor within the class DIII of symmetries.


However, the introduction of a vortex degrades the symmetry down to class D by breaking time-reversal, and can lead to surface Majorana zero modes (MZMs) for a finite range of parameters.\cite{fu2008superconducting,hosur2011majorana} For our purposes, we have followed Ref. \onlinecite{hosur2011majorana} and introduced a vortex line defect in 3D by hand. More precisely, we let the $\Delta(\mathbf{r})$ vary in real space according to the following exponentially relaxing form
\begin{align}
\Delta(\mathbf{r}) = \Delta_0\, \mathrm{e}^{i \phi(\mathbf{r})}\, \left[1-e^{-\rho(\mathbf{r})/\xi}\right]
\label{eqn:vortex_delta}
\end{align}
with $\mathbf{r}=(x,y,z)$, $\tan \phi(\mathbf{r}) = (y-y_0)/(x-x_0)$ and $\rho(\mathbf{r})^2 =\nobreak (x-x_0)^2+(y-y_0)^2$. The nodal line is situated at $(x,y)=(x_0,y_0)$. The SC coherence length is $\xi$ and $\Delta_0$ is the bulk SC order parameter strength.


Shown in Fig.~\ref{fig:MZM}(a) is an energy spectrum of the model with a vortex line. It contains a pair of Majorana zero modes (MZMs) localized on the top and bottom surfaces where the vortex penetrates and leaves the bulk SC. The results are obtained from numerical Lanczos exact diagonalization of a large $40\times 40 \times 40$ crystal with open boundary conditions. The MZMs are stable topological boundary modes for a finite range of parameter but become unstable when the chemical potential is sufficiently deep inside the conduction band. This proceeds as a topological transition within the vortex line through the proliferation of gapless Carroli-de Gennes-Matricon modes into the bulk.\cite{hosur2011majorana} Due to the symmetries of the problem, the vortex line defect can best be understood as a topological ``Kitaev wire'' embedded within a bulk trivial SC. 


\begin{figure}
\includegraphics[width=5cm]{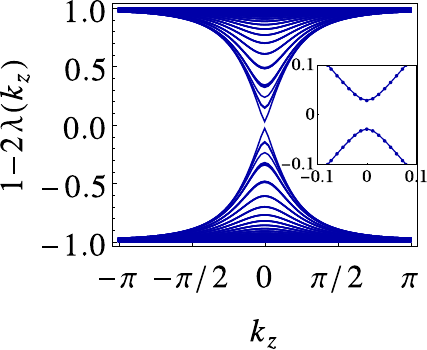}
\caption{Entanglement spectrum from isolating a $9\times 9$ square region centered on the vortex nodal line in a $40 \times 40$ sized SC. The entanglement spectrum is gapped as is emphasized by the inset which is a blow up of the small gap region with individual data points highlighted. Calculation of the Zak-Berry phase $\theta_B$ from the `occupied' entanglement eigenstates gives a value of $\pi$ indicating a topologically non-trivial superconducting ETI wire in class D.}
\label{fig:ETI_V_ES}
\end{figure}

Next we apply the disentangling method to validate our interpretation of the SC vortex as being a one dimensional ETI in class D. To do so, we consider a $40 \times 40$ square that is infinite and translationally symmetric in the $+z$ direction but with open boundaries in the $xy$-plane. The nodal line of the vortex is placed at the center and with the same model parameters as in Figs. \ref{fig:ETI_V} and \ref{fig:MZM}. Shown in Fig.~\ref{fig:ETI_V_ES} is the resulting entanglement spectrum obtained by tracing over all sites except a $9 \times 9$ square region (in the $xy$-plane) centered at the nodal line. It shows a gapped entanglement spectrum with a minimal gap at $k_z=0$ which is quite small but still present as is highlighted by the accompanying inset. 

The class D of symmetries in 1D are classified by a $\mathbb{Z}_2$ index. Topologically non-trivial class-D wires can be detected through the sign of the Pfaffian associated to the antisymmetric matrix representation of the Hamiltonian written in the Majorana operator basis.\cite{kitaev2001unpaired} Equivalent to the Pfaffian $\mathbb{Z}_2$ index is the Zak-Berry phase (modulo $2\pi$) or polarization\cite{budich2013equivalent} which is more straightforward to compute. Specifically, the Zak-Berry phase or Entanglement Berry phase\cite{fukui15_disen} as it is known in this context has the expression
\begin{align}
\theta_B = \int_{-\pi}^\pi \mathrm{d}k_z \; \text{tr} \,\mathcal{A}_z(k_z) 
\end{align}
where 
\begin{align}
[\mathcal{A}_z(k_z)]_{mn} = \langle \lambda^{\mathcal{(R)}}_m(k_z) | i \nabla_{k_z} \lambda^{(\mathcal{R})}_n(k_z) \rangle
\end{align}
is the non-Abelian Berry connection derived from the `occupied' entanglement eigenstates $\{|\lambda_n\rangle : \lambda_n > 1/2\}$ of the restricted projector $\rho_\mathcal{R}$. Here $\mathcal{R}$ is the square region that contains the vortex nodal line. For a trivial wire, $\theta_B = 0$ mod $2\pi$ and when non-trivial $\theta_B = \pi$ mod $2\pi$. For practical purposes, it is best to carry out the computation of $\theta_B$ using a discretized grid of $k_z$ points and the discretized gauge invariant Wilson loop expression 
\begin{subequations}
\begin{align}
\theta_B &= \text{Im}\,\log \prod_{i=1}^{N_\text{grid}} \det \mathcal{U}_i \\ 
[\mathcal{U}_i]_{mn} &= \langle \lambda^{(\mathcal{R})}_m (k_i)|\lambda^{(\mathcal{R})}_n(k_{i+1})\rangle.
\end{align} 
\end{subequations}
For the data shown in Fig.~\ref{fig:ETI_V_ES}, a uniform grid of 80 points yields a value of $\theta_B =\pi$ confirming the presence of an ETI in agreement with the observance of Majorana zero modes in Fig.~\ref{fig:MZM}.

\section{III. Two Embedded Topological Insulators and Quantum Mutual Information}
\label{supp:two_ETIs}

In this supplemental section we consider in more detail the effect of having two finitely separated ETIs, and the net effect this has on surface states. This is then quantified in the bulk by the use of the quantum mutual information. 

\subsection{IIIA. Mutual Information} 
This is a measure based on the entanglement entropy $S_A$ which has the expression\cite{peschel2003calculation}
\begin{align*}
S_A = -\sum_{i}\left[
\lambda^{(A)}_i \log_2 \lambda^{(A)}_i+(1-\lambda^{(A)}_i) \log_2(1-\lambda^{(A)}_i)
\right]
\end{align*}
for a region $A$ and specialized to the case of free fermions. Here $\{\lambda_i^{(A)}\}$ are the set of eigenvalues of the 1-body reduced density matrix $\rho_A$ in region $A$. The mutual information is given as
\begin{align}
I_2[A;B] = S_A + S_B - S_{A\cup B}
\end{align}
and acts as a bound for connected correlation functions involving operators in $A$ and $B$.
Specifically, the following connected correlation function is bounded\cite{wolf2008area} as
\begin{align}
&\langle \mathscr{A} \otimes \mathscr{B}\rangle_c^2 \leq 2 \Vert \mathscr{A} \Vert^2 \Vert \mathscr{B} \Vert^2 I_2(A:B) \end{align}
where $\mathscr{A}$ and $\mathscr{B}$ are any bounded (many-body) operators with support in $A$ and $B$ respectively, and with operators norms $\Vert \mathscr{A} \Vert,\Vert \mathscr{B} \Vert$. Specializing to Gaussian states and the 1-body fermionic correlator yields
\begin{align}
|\langle c^\dagger_\alpha c_\beta \rangle|^4 \leq 2 I_2 (A:B)
\label{eqn:hopping_bound}
\end{align}
where $\alpha,\beta$ are orbitals located in $A$ and $B$ respectively such that $\{c_\alpha, c^\dagger_\beta\} = \delta_{\alpha \beta}$.
Whenever the momentum $\mathbf{k}$ is a good quantum number, the entanglement entropies and mutual information $I_2[A;B]$ may also be $\mathbf{k}$-resolved. The latter yields a `mutual information band structure' $I_2[A;B](\mathbf{k})$.
In practice, we will take $A$ and $B$ to be two different ETIs embedded in the same environment where $I_2[A;B]$ will track the amount of `hybridization' between $A$ and $B$. A small amount of mutual information between them would imply relatively weak coupling and hence relatively weakly coupled surface states with small energy gaps if present.


\begin{figure}
\includegraphics[width=4.6cm]{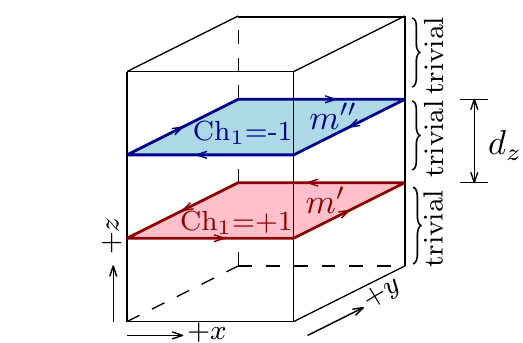}
\caption{(color online) Schematic of the two embedded Chern Insulators (ECIs). The lower (red) and upper (blue) impurity layers are ECIs of opposite topological invariants Ch$_1=\pm 1$ and $d_z$ denotes the number of trivial layers between them.}
\label{fig:ETI_CI_double}
\end{figure}

\begin{figure}
\includegraphics[width=4.2cm]{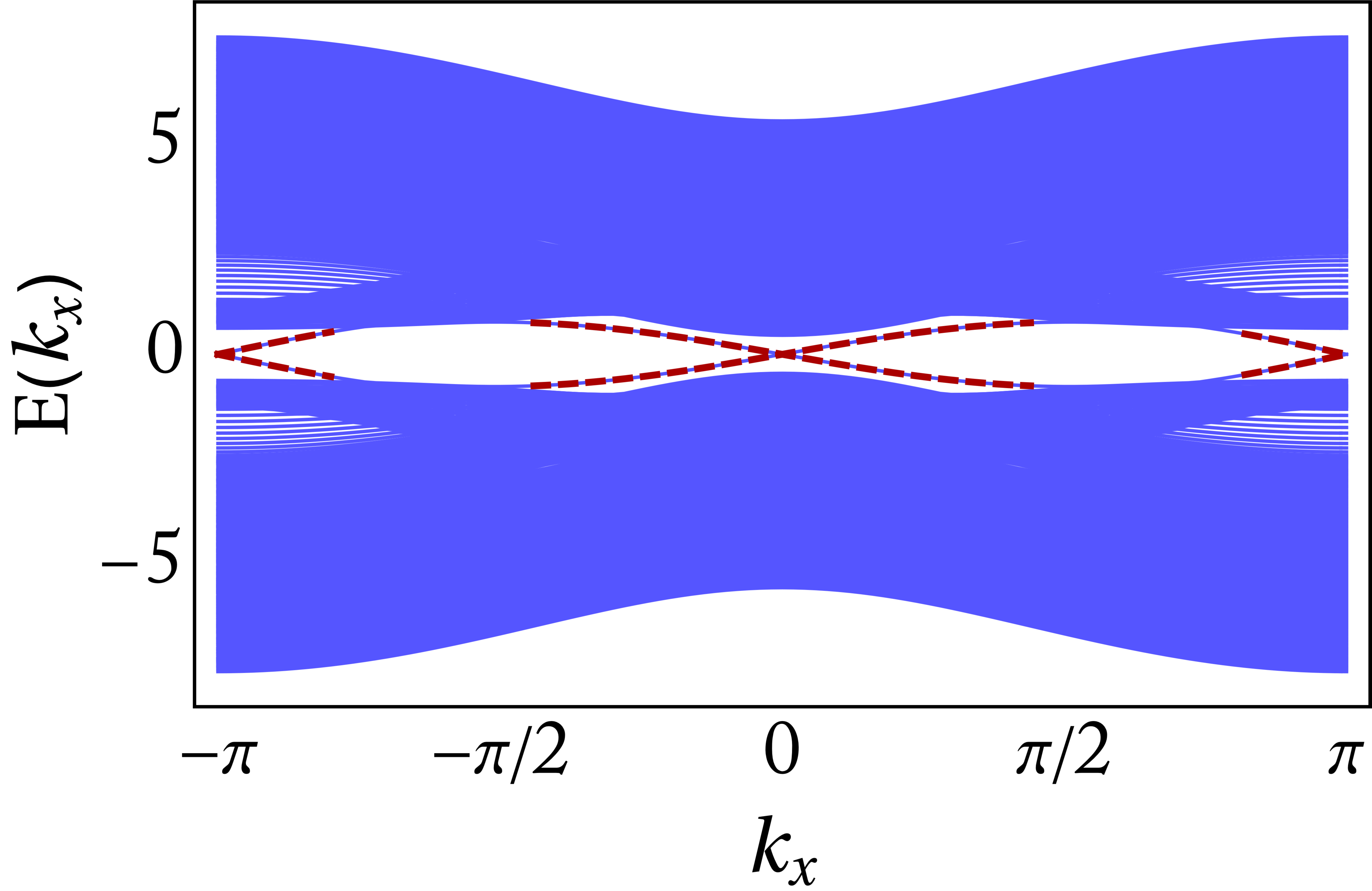}
\caption{(color online) Energy band structure of two ECIs (Fig.~\ref{fig:ETI_CI_double}) with $d_z =4$ within a crystal of finite dimensions $L_y=L_z=20$ and has PBC in $z$ but OBC in $y$. Parameters are $(m,m',m'',\delta_z)=(-2,1,3,0.8)$ such that the two ECIs have opposite Chern numbers. Dashed (red) lines are gapless topological edge modes that cross at $k_x=0,\pi$ and localize on the different CI layers.}
\label{fig:ETI_CI_BS_double_top}
\end{figure}

\subsection{IIIB. Two Embedded Chern Insulators} 

Now we consider a situation where there are two closely located Embedded Chern Insulators (ECIs). A schematic of the composite system is shown in Fig.~\ref{fig:ETI_CI_double} where the two CIs are taken to have opposite Chern numbers Ch$_1=\pm 1$. The model that we use is the one in Supplementary Section IIB with some modifications. This can be done in two different ways leading to different types of surface modes. But viewed as an entire 2D band insulator (in $xy$), the composite system is classified as being topologically trivial (Ch$_1=-1+1=0$). Nevertheless, surface states are present like in previous example and in the large $d_z$ limit are effectively independent topological chiral edge modes.

\begin{figure}
\includegraphics[width=0.48\textwidth]{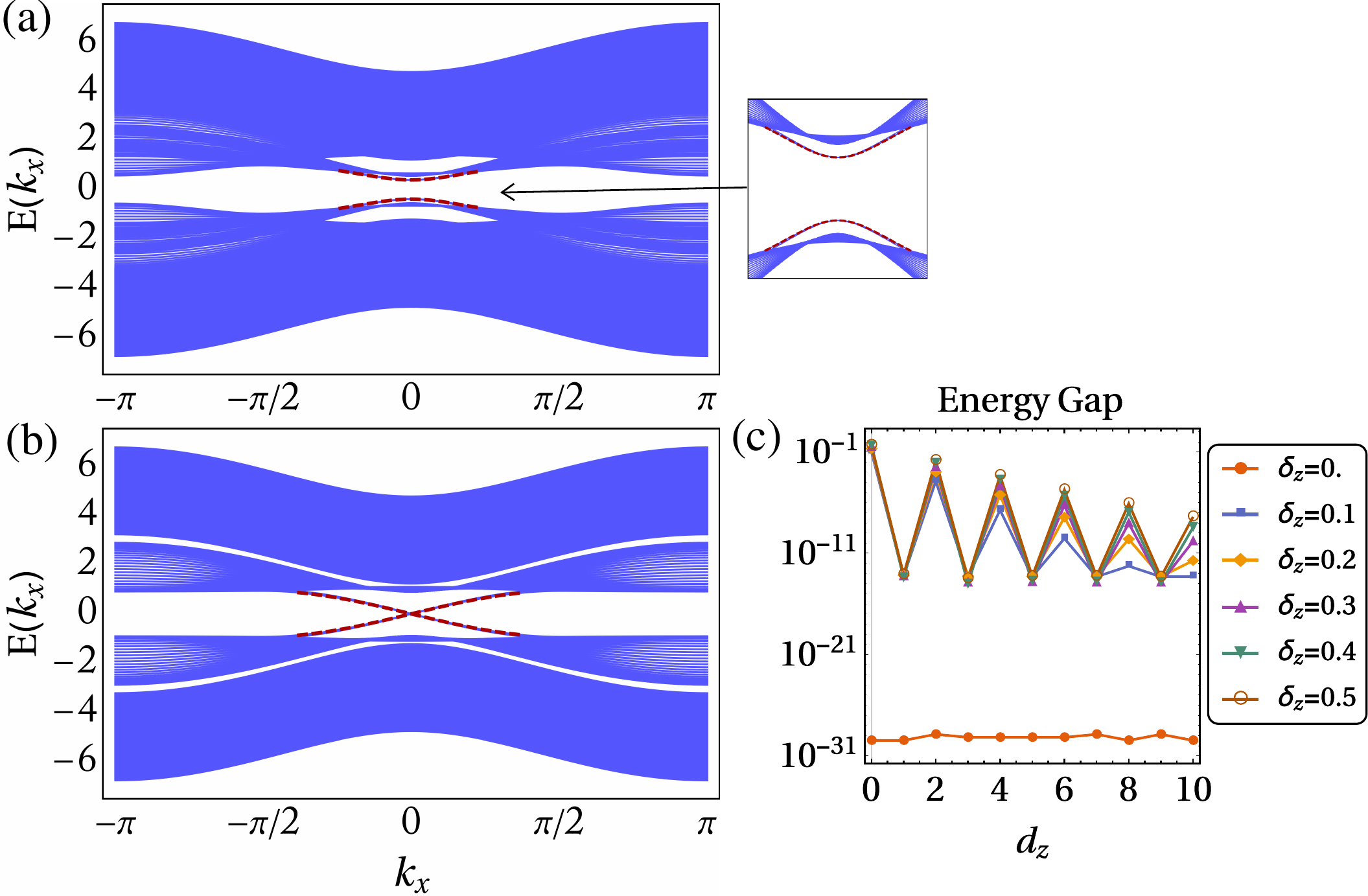}
\caption{(color online)
Energy band structure data of two ECIs (Fig.~\ref{fig:ETI_CI_double}) with parameters $(m,m',m'')= (-2,1,1)$ at varying distance $d_z$ and interlayer coupling $\delta_z$. The CIs are taken to be time-reversed partners of each other giving opposite Chern numbers, and edge states (red dashed lines) centered at $k_x=0$. The system has finite dimensions $L_y=L_z=20$ with PBC in $z$ but OBC in $y$.
(a) Gapped edge states when $d_z=0$ and $\delta_z=0.4$.
(b) Near gapless edge states when $d_z=8$ and $\delta_z=0.4$.
(c) Energy gaps at $k_x=0$ on a log scale for varying $d_z$ and $\delta_z$ that are finite size effect and machine precision limited.
}\label{fig:ETI_CI_double_dz_composite}
\end{figure}

First, we denote the two mass parameters for the different CI layers in Fig.~\ref{fig:ETI_CI_double} as $m'$ for the bottom (red) and $m''$ for the top (blue) layers respectively. Taking $0<m'<2$ and $2<m''<4$, with $d_z \neq 0$  leads to the energy band structure shown in Fig.~\ref{fig:ETI_CI_BS_double_top} with open $y$ boundaries. There are oppositely dispersing gapless chiral edges centered at $k_x=0,\pi$. Moreover, these modes are localized on the different impurity layers as expected and remain gapless independent of distance $d_z$ so long as the bulk energy gap remains open.

A second way to obtain a Ch$_1=-1$ embedded Chern Insulator is to use a time reversed partner of the Ch$_1=+1$ model. This leads to the energy band structure of Fig.~\ref{fig:ETI_CI_double_dz_composite} with surface states centered around $k_x=0$ only. In this instance, an energy gap develops in the surface states (Fig.~\ref{fig:ETI_CI_double_dz_composite}(a)) that rapidly decays with increasing distance (Fig.~\ref{fig:ETI_CI_double_dz_composite}(b,c)). We note that there is an even-odd effect with respect to $d_z$, where for odd finite $d_z$ the energy gap goes to zero in the infinite $L_y$ limit.


\begin{figure}
\includegraphics[width=0.4\textwidth]{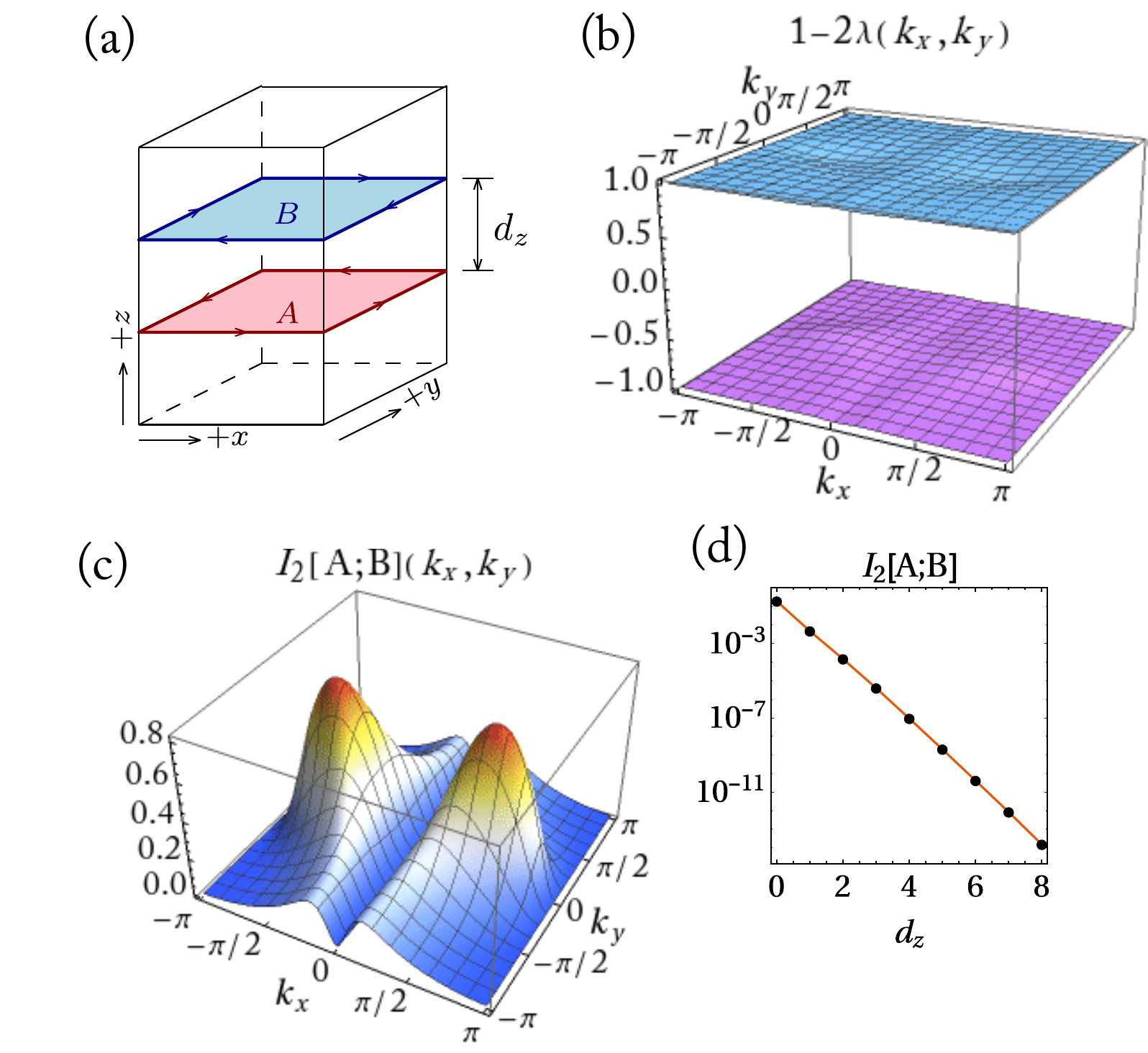}
\caption{(color online) Mutual information measures of the two ECIs related by time-reversal. Parameters are $(m,m',m'',\delta_z)=(-2,1,1,0.4)$ with $L_z=24$. (a) Schematic of the two ECIs labeled here as regions $A$ and $B$. (b) The gapped entanglement spectrum band structure from restriction to the $A$ ECI with $d_z=0$. The lower band has Chern number Ch$_1=+1$. (c) Mutual information $I_2[A;B](\mathbf{k})$ band structure between the two ECIs at $d_z=0$. (d) The mutual information $I_2[A;B]$ averaged over the Brillouin zone at varying $d_z$.}
\label{fig:ETI_CI_ES_double_dz_composite}
\end{figure}

\begin{figure}
\includegraphics[width=0.45\textwidth]{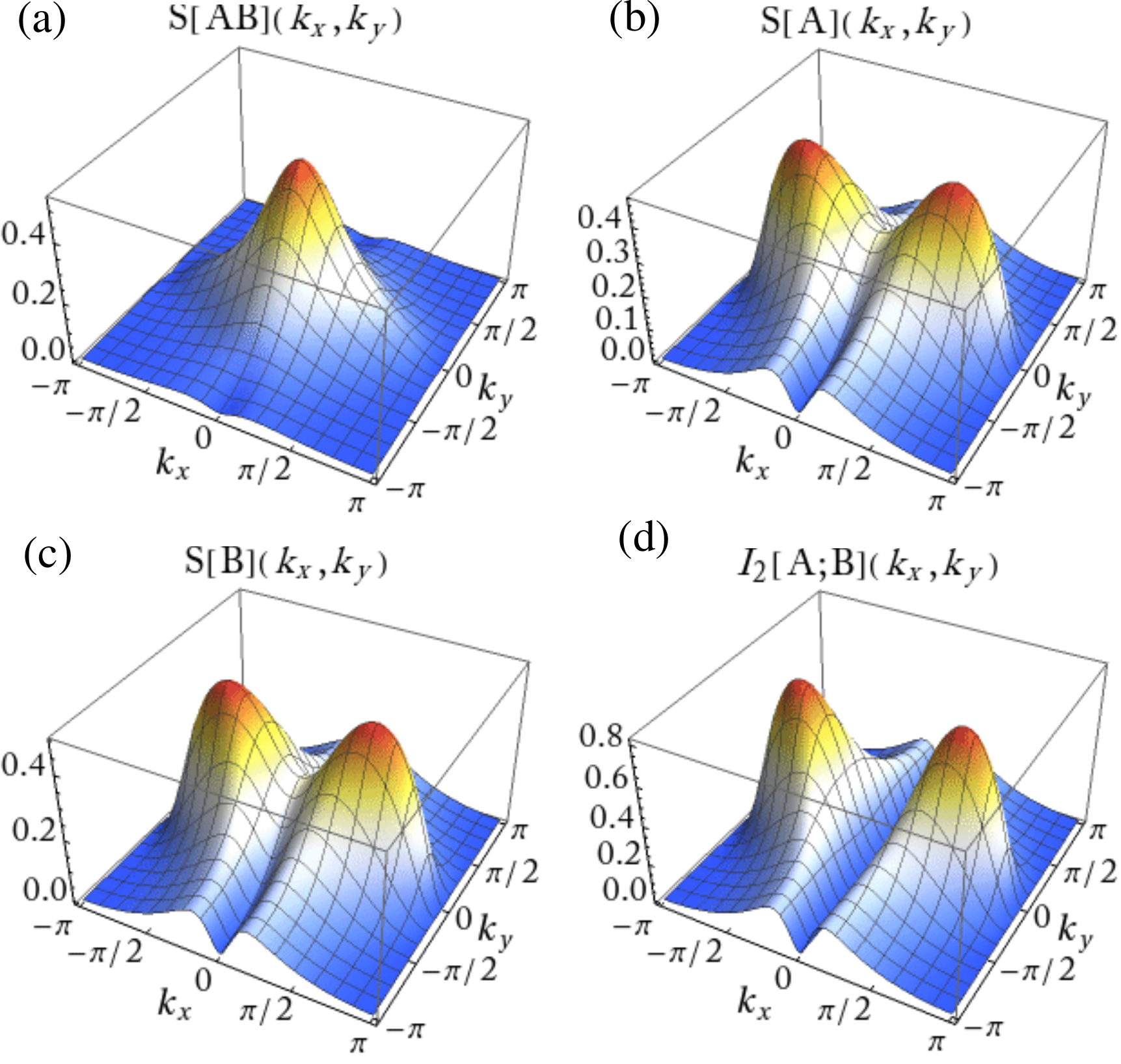}
\caption{The breakdown of the different entanglement entropy bands (a) $S[AB](\mathbf{k})$, (b) $S[A](\mathbf{k})$, (c) $S[B](\mathbf{k})$ that make up the mutual information band structure $I_2[A;B](\mathbf{k})= S[A](\mathbf{k})+S[B](\mathbf{k})-S[AB](\mathbf{k})$.}
\label{fig:I2_all}
\end{figure}

Next, shown in Fig.~\ref{fig:ETI_CI_ES_double_dz_composite}(b) is the entanglement spectrum band structure for a single ECI layer in the case that the two ECIs are adjacent with $d_z=0$. The associated mutual information band structure between ECIs is shown in Fig.~\ref{fig:ETI_CI_ES_double_dz_composite}(c). Note the maxima in $I_2[A;B](\mathbf{k})$ near $k_x = \pm\pi/2$ that correspond to the emergence of the surface states from the bulk [See Fig~\ref{fig:ETI_CI_double_dz_composite}(b)]. Crucially, the entanglement spectrum bands are gapped with quantized Chern number Ch$_1=+ 1$ even when $d_z=0$, thus qualifying the ECI status. However as is evident in Fig.~\ref{fig:ETI_CI_double_dz_composite}(a), the edge state spectrum is gapped meaning that the surface states do not enjoy any topological protection because of edge mode coupling. This perfectly illustrates the situation that non-trivial quantized topological invariants derived from entanglement bands are \emph{no guarantee} for the integrity of the topologically protected surface modes. Physically, this is because the surface states may still be gapped without closing the bulk gap if there are available surface states from compensating ETIs. Only in the case of sufficiently separated ETIs do we expect to recover non-trivial topological surface states as exemplified in Fig.~\ref{fig:ETI_CI_double_dz_composite}(b,c). This picture is further supported by the decay in mutual information between ECIs as shown in Fig.~\ref{fig:ETI_CI_ES_double_dz_composite}(d). {However, it is interesting to note that the exponential decay length of the surface energy gap and the averaged mutual information differ by about a factor of 2 with the energy gap decaying slower. We attribute this to the fact that surface states as being more delocalized than the bulk states, of which the mutual information is only sensitive to the latter. Nevertheless the mutual information -- as computed in this way -- may still serve as a useful quantitative guide for the amount of coupling between ETIs utilizing only bulk information.} Finally, for the sake of completeness, the individual contributions that go into the mutual information band structure $I_2[A;B]$ of Fig.~\ref{fig:ETI_CI_ES_double_dz_composite}(c) is shown in Fig.~\ref{fig:I2_all}.

\section{IV. Stacking Faults and Partial Dislocations}
\label{supp:defects}

The stacking fault and partial dislocation defects models in Fig.~\ref{fig:stacking} are derived from a pristine ETI crystal of stacked CIs; termed $A$ and $B$ type. The ETI crystal has the Bloch Hamiltonian
\begin{align*}
H(\mathbf{k}) & = 
\sin k_x (\sigma^z \otimes \sigma^x) +
\sin k_y (\mathbbm{1} \otimes \sigma^y) \nonumber \\
& +[2-m_1-\cos k_x - \cos k_y] (\mathbbm{1} \otimes \sigma^z) \nonumber \\
& + [\gamma_z + \delta_z \cos k_z] (\sigma^x \otimes \mathbbm{1}) 
- m_2 (\sigma^z \otimes \sigma^z)
\end{align*}
where $m_1,m_2,\gamma_z$ and $\delta_z$ are model parameters. Physically, each CI subsystem within the unit cell is modeled on the CIs given in Section IIB or Eqn.~(\ref{eqn:CI_Ham}) but which are `time-reversed' relative to one and other. Moreover, we include differences to their on-site $m$ parameters. Namely, $m_1$ is their average $m$-parameter while $m_2$ quantifies the their difference. Lastly, `inter CI' hopping parameters are quantified by $\gamma_z$ and $\delta_z$ which are the intra-unit cell and inter-unit cell respectively.

From this Bloch-Hamiltonian, the real-space form of the Hamiltonian can be obtained by an inverse Fourier transform, performed only on the $y,z$ directions which are no-longer periodic in the presence of the defects. The defected systems are derived by altering this hybrid real-space ($y,z$) and momentum space ($k_x$) pristine ETI crystal model Hamiltonian. Conventionally, we take the periodic boundary conditions in the $+z$ direction but open boundary conditions in $+y$ for the stacking fault, and periodic boundary conditions in $+y$ for the partial dislocation. Thus only the stacking fault has an exposed $yz$ surface.
The stacking fault is created in real-space
by the removal a $B$ layer and rejoining the resulting exposed $A$ layers,
\ie the Volterra construction.
The stacking fault spectrum in Fig.~\ref{fig:stacking}(d)
was created with the parameters
$m_1=1$, $m_2=-0.1$, $\gamma_z=0.05$, and $\delta_z=0.5$.
The partial dislocation is created
by truncating half of a $B$ layer in the $y$-direction
and rejoining the now-exposed portion of the $A$ layers.
The partial dislocation spectrum in Fig.~\ref{fig:stacking}(f)
was created with the parameters
$m_1=1$, $m_2=0$, $\gamma_z=0.5$, and $\delta_z=0.25$.
The topological edge states
remain robust to different model parameter values
and the manner in which the rejoining operation is performed.

\section{V. Classification of Topological Insulators by Spectral Projectors}
\label{supp:P}

In this supplemental section, we describe for the benefit of readers unfamiliar with spectral projectors, the calculation of topological indices or invariants using $P$. Beginning with the gapped Bloch Hamiltonian $H(\kk)$ for translationally invariant insulators, $P(\kk)$ can be expressed as
\begin{align*}
P(\kk) = \oint_\gamma \frac{\dd z}{i 2\pi} \; \frac{1}{z + E_F - H(\kk)}
\end{align*}
where the contour $\gamma$ encloses the negative part of the $\mathbb{R}$ line in an anti-clockwise fashion. Assuming that $H(\kk)$ is analytic and a modified form of the Paley-Wiener theorem,\cite{kuchment2016overview} leads to the decay of $P$ in real space
\begin{align*}
P(\RR,\RR') &= \int_\text{BZ} \frac{\dd^d k}{|\mathcal{C}^*|} P(\kk) \e^{i\kk \cdot(\RR-\RR')} \nonumber \\
|P(\RR,\RR')| &< A \e\left( \left[\tfrac{|\RR-\RR'|}{\xi}\right]^d\right)
\end{align*}
as $|\RR-\RR'| \rightarrow \infty$  for some  $A,\xi > 0$. Here $\RR$ and $\RR'$ are lattice site positions. This is just the well known statement that non-interacting band gap insulators have exponentially bounded correlations.

Now the way that $P$ defines a vector bundle $\mathbb{V}_\text{occ}$ of occupied Bloch states over the BZ is expressed as
\begin{align*}
v \in \mathbb{V}_\text{occ} \Leftrightarrow P(\kk) |v(\kk)\rangle = |v(\kk)\rangle \quad \forall\; \kk \in \text{BZ}.
\end{align*}
Depending on the symmetry and dimension of the BZ, the computation of the relevant topological indices,\cite{chiu2016classification} which we have denoted by $c([P])$ in the main text, is carried out either with $P$ directly or by selecting a frame bundle from $P$ with a specific gauge choice. An example of the former case is the 2D Chern number in symmetry class A that has the following gauge invariant expression
\begin{align*}
\text{Ch}_1 = \frac{i}{2\pi} \int_\text{BZ} \;  \text{tr} \left[ P(\kk) \;dP(\kk) \wedge dP(\kk) \right] \in \mathbb{Z}.
\end{align*}
A second example which uses a gauge choice of Bloch basis for convenience is the computation of the strong index for a time-reversal invariant insulator in 3D in class AII. This requires selecting a globally smooth orthonormal basis of Bloch states $\{u_1,u_2,\ldots u_m \}$ that span $ \mathbb{V}_\text{occ}$, which can always be done because of time-reversal symmetry. Using the sewing matrix which has the expression
\begin{align*}
w_{\alpha \beta}(\kk) = \langle u_\alpha (-\kk) | T u_\beta(\kk) \rangle
\end{align*}
where $T$ is the time-reversal operator squaring to (-1), the strong index $\mathbb{Z}_2$ is then given as
\begin{align*}
\nu = \prod_{\K_i} \frac{\text{Pf} \;w(\K_i)}{\sqrt{\text{det} \;w(\K_i)}}
\end{align*}
where the product extends over all time-reversal invariant momenta. Alternatively the strong index may be computed from the non-gauge invariant Chern-Simons integral derived from the specific choice of $\{u_n\}$. Also for the chiral classes in odd dimensions, decomposing $P$ into chiral eigenspaces can be used to compute winding numbers that are the topological invariants of that class.

\section{VI. Strong Topological Indices of Chiral Symmetric Wires}
\label{supp:nu_cal}


In this appendix section, we describe computational aspects related to the determination of the strong topological index $\nu \in \mathbb{Z}$ that classifies chiral symmetric wires. But first we clarify some subtleties regarding the choice of plane-wave basis, unit cells and the $k-$space connection which are involved in computing $\nu$.

There are two popular conventions for the plane-wave basis used to define the periodic part of the Bloch wavefunction. Moreover, the chiral symmetric topological classes (AIII,BDI,CII) are sensitive to the particular choice of basis for reasons that shall explain below. The most natural choice of basis follows from the continuum definition adapted to the tight-binding lattice and is given by the following infinite sum of real-space kets
\begin{align}
|\kk, \rr_\alpha \rangle = \sum_{\RR \in \Lambda} \e^{i \kk \cdot (\RR + \rr_\alpha)} | \RR + \rr_\alpha \rangle
\end{align}
where $\Lambda$ is the Bravais lattice in $d$-dimensions, $\rr_\alpha$ is the position of the $\alpha$-th tight-binding orbital within the unit cell and $\alpha$ varies over the different orbitals. Bloch states are then expressed as
\begin{align}
|\Psi_{\kk}\rangle = \int_\text{BZ}\frac{\diff^d k}{|\mathcal{C}^*|} \sum_{\alpha} u_{\alpha}(\kk)|\kk,\rr_\alpha\rangle
\end{align}
with $|\mathcal{C}^*|$ being the volume of the 1st Brillouin zone (BZ). This choice is natural because the plane-wave states $|\kk,\rr_\alpha\rangle$ transform as a representation of the Euclidean group $\mathbb{E}^d$ as
\begin{align}
|\kk,\rr_\alpha\rangle \rightarrow
 |M\cdot\kk,M\cdot\rr_\alpha + {\bf a}\rangle, \quad
M\in \text{O}(d), \; {\bf a}\in \mathbb{R}^d.
\end{align}
This is because the basis is conscious of the orbital locations within the unit cell and hence transforms as a representation space group of the crystal lattice. Moreover, the coefficients of the Bloch eigenstates $u_\alpha(\kk)$ are independent of the origin of the unit cell in this basis. This is due to the fact that $H(k)$ only cares about hopping distances between orbital sites in this basis. However the basis is not periodic in the BZ with $|\kk + \GG, \rr_\alpha\rangle = \e^{i \GG\cdot \rr_\alpha} |\kk,\rr_\alpha\rangle$ that requires $u_\alpha(\kk+\GG)=\e^{-i \GG\cdot \rr_\alpha} u_\alpha(\kk)$, where $\GG \in \Lambda^*$ is a reciprocal lattice vector.

The other competing choice of Bloch basis is
\begin{align}
\widetilde{|\kk,\rr_\alpha\rangle} =\sum_{\RR \in \Lambda} \e^{i \kk \cdot \RR} | \RR + \rr_\alpha \rangle =  \e^{-i \kk \cdot \rr_\alpha}|\kk,\rr_\alpha\rangle
\end{align}
which is BZ periodic but has less appealing transformation properties. In this case, information regarding the orbital positions have been transformed away. In fact, the two competing bases are related by the non-regular gauge transformation $\widetilde{|\kk,\rr_\alpha\rangle} = \e^{-i \kk \cdot \rr_\alpha}|\kk,\rr_\alpha\rangle$, in which the phase function is not BZ periodic.

The reason this matters is because the intra unit cell orbital configuration is absolutely crucial to understanding and classifying the chiral symmetric class. For example, in the Su-Schrieffer-Heeger (SSH) chain\cite{su1979solitons} -- which is in the BDI class -- the choice of unit cell that bisects a dimerized bond leads to a topologically non-trivial chain. Whereas, the complementary choice of unit cell which includes whole dimers is trivial. Hence, the choice of unit cell needs to be reflected in the $k-$space covariant derivative $\nabla_\kk$ which is used to determine $\nu$. The definition of the covariant derivative that is appropriate turns out to be
\begin{align*}
\nabla_\kk = \partial_\kk + i\,  \hat{\rr}
\end{align*}
and is such that the Berry connection computed with either choice of basis is identical. Note that $\hat{\rr}$ is the orbital position operator such that $\hat{\rr}|\kk,\rr_\alpha\rangle = \rr_\alpha |\kk,\rr_\alpha\rangle$ and is zero in the $\widetilde{|\kk,\rr_\alpha\rangle}$ basis. Also,  $i\nabla_\kk$ is just the unit cell position operator $\hat{\RR}$ expressed in the plane-wave basis.

Now, the chiral topological insulator classes (AIII,BDI,CII) in 1D are classified by the topological index $\nu$ that has the expression as a chiral (skew) polarization quantity~\cite{ryu_topological_2010,mondragon-shem_topological_2014}
\begin{align}
\nu = \frac{1}{\pi}\int {\diff k}\sum_{n}
\langle u_{n}(k)| i\,S\,\nabla_k u_n(k)\rangle \in \mathbb{Z}
\end{align}
where $|u_n(k)\rangle = \sum_{\alpha} u_{n,\alpha}(k) |\alpha\rangle$ are the Bloch eigenstates and the $n$ sum extends over occupied $E_n(k)$ bands. The operator $S$ acts on orbital degrees of freedom and is the chiral grading operator that anti-commutes with the Bloch Hamiltonian. Because of the identification of $i\nabla_{k}$ with $\hat{R}$, this expression of $\nu$ has the interpretation of a 1D polarization weighted by $S$.

However there is an alternative interpretation of $\nu$ that is more useful in revealing the origin of its quantization. First a re-writing the Bloch Hamiltonian into blocks of definite $S$ chirality yields the following block off-diagonal form
\begin{align}
H(k) = \sum_{n=1}^N E_n(k) \begin{pmatrix}
0_N & q_n(k)^\dagger \\ q_n(k) & 0_N
\end{pmatrix}
\end{align}
for a system of $2N$ energy bands. Here the first $N$ orbital states are $S=+1$ definite while the remaining $N$ are $S=-1$, and $0_N$ is the $N\times N$ zero matrix. The sub-matrices $q_n(k)$ are defined as
\begin{align}
[q_n(k)]_{\alpha \dot{\alpha}} = [u_{n\uparrow}(k)]_\alpha[u_{n\downarrow}(k)]_{\dot\alpha}^*,\quad  \alpha,\dot\alpha=1,\ldots N
\end{align}
where
\begin{align}
u_{n\uparrow}(k) &= \tfrac{1}{\sqrt{2}}\left( u_n(k) +S u_n(k) \right) \\
u_{n\downarrow}(k) &= \tfrac{1}{\sqrt{2}}\left( u_n(k) -S u_n(k) \right)
\end{align}
are projections of the Bloch eigenstates $u_n(k)$ with positive energy $E_n(k)>0$ onto definite $S$ chirality. A gapped $H(k)$ guarantees that there is always a well defined separation between the negative and positive eigenstates and hence a smooth $q_n(k)$. It can be shown that the chiral winding is equivalently expressed as
\begin{align}
\nu = \frac{1}{2\pi}\int \diff k \, \text{Tr}\left[
Q(k)^\dagger i \nabla_{k} Q(k)
\right]
\label{eqn:nu}
\end{align}
where $Q(k) = \sum_{n=1}^N q_n(k) \in \text{U}(N)$ is unitary. Now the use of the covariant derivative $\nabla_k = \partial_k + i \hat{r}$ is absolutely necessary because the matrix $Q(k)$ is not BZ periodic. Nevertheless by transforming to the $\widetilde{|k,r_\alpha\rangle}$ basis with the transformation
\begin{align*}
\tilde{Q}(k) = D(k)^\dagger Q(k)D(k)
\end{align*}
with $D(k)= \e^{ik \, \hat{r}}$ yields a $\tilde{Q}(k)$ that is unitary and BZ periodic. Then $\nu$ is
\begin{align}
\nu = \frac{1}{2\pi}\int \diff k \, \text{Tr}\left[
\tilde{Q}(k)^\dagger i \partial_{k} \tilde{Q}(k)
\right]
\end{align}
which measures to the winding of the phase $\det\, \tilde{Q}(k) \in \text{U}(1)$ around the 1D BZ. It is manifestly quantized due to the BZ periodicity of $\tilde{Q}(k)$. In practical numerical experiments, the determination of $\nu$ proceeds not with computing the integral in (\ref{eqn:nu}) but by plotting in the phase of $\det\, \tilde{Q}(k)$ as a function of $k$ to determine the number of times it winds around zero.

\end{document}